# Theory of terahertz pulse transmission through ferroelectric nanomembranes

Yujie Zhu[1], Aiden Ross[2], Xiangwei Guo[1], Venkatraman Gopalan[2], Long-Qing Chen[2], and Jia-Mian Hu[1*]

[1]*Department of Materials Science and Engineering, University of Wisconsin-Madison, Madison, WI, 53706, USA*

[2]*Department of Materials Science and Engineering, The Pennsylvania State University, University Park, Pennsylvania, 16802, USA*

## Abstract

An analytical model is developed to predict the temporal evolution of the lattice polarization in ferroelectric nanomembranes upon the excitation by a terahertz (THz) electromagnetic pulse of an arbitrary waveform, and the concurrent transmission of the THz pulse in both the linear and the nonlinear regimes. It involves the use of the perturbation method to solve the equation of motion for the lattice polarization in both unclamped and strained ferroelectric nanomembranes within the framework of Landau-Ginzburg-Devonshire theory. The model is applicable to perovskite oxides such as $BaTiO_3$ and $SrTiO_3$, wurtzite $Al_{1-x}Sc_xN$, and trigonal $LiNbO_3$. Our analytical model provides a theoretical basis for determining the thermodynamic and kinetic parameters of ferroelectric materials through THz transmission experiment. The calculation results also suggest an approach to reversing the chirality of a circularly polarized THz pulse by harnessing the resonant polarization-photon coupling in ferroelectrics. This capability of chirality reversal, along with the high tunability from a strain applied along any arbitrarily oriented in-plane axis, provides new opportunities for THz wave modulation without relying on complex metasurface designs.

[*]E-mail: jhu238@wisc.edu

## I. Introduction

The elementary excitations of the charge, spin, orbital, and lattice degrees of freedom in many functional and quantum materials have an energy scale within the terahertz (THz) regime [1]. Therefore, THz radiation can probe, excite, and control these excitations to realize emergent phenomena [2,3]. Examples include the creation of a THz-field-induced non-equilibrium superconducting-like state in $YBa_2Cu_3O_{6.5}$ [4], inducing an ultrafast martensitic phase transition in a $Nb_3Sb$ thin film [5], a shift in the Fermi surface in the topological material $Bi_2Te_3$ [6], the insulator-to-metal transition in $VO_2$-based metamaterials [7], a non-thermal spin switching in antiferromagnetic $Sm_{0.7}Er_{0.3}FeO_3$ [8], and the excitation of a magnetoelectric oscillation in multiferroic $TbMnO_3$ [9].

In ferroelectric materials, intense THz pulses can drive the soft mode phonon to the anharmonic regime [10–15], enabling the evaluation of the high-order coefficients in the interatomic potential [10–12,14] and facilitating the energy transfer to higher-frequency phonon modes through nonlinear coupling [12]. Moreover, they have been used to rotate the lattice polarization [16–18], induce nonlinear optical phenomena [19,20], excite collective motion of topological polar textures [21], drive transient paraelectric-to-ferroelectric phase transitions [22], and create non-equilibrium polar phases [23,24]. Furthermore, circularly polarized THz pulses have been employed to drive the chiral motion of lattice polarization and the resulting emergence of an effective magnetization in $SrTiO_3$ bulk single crystals (namely, dynamical multiferroicity) [8].

The main objective of this paper is to establish an analytical model to understand and predict the interaction between a THz pulse of an arbitrary waveform and a ferroelectric nanomembrane, which necessitates the consideration of the coupled dynamics among the lattice polarization (soft mode), strain, and electromagnetic waves (photons). This is because the lattice polarization and strain are mutually coupled via the piezoelectric coupling even at the picosecond (ps) scale [25,26], and the polarization-photon coupling, i.e., the influence of the polarization current (the $\partial\mathbf{P}/\partial t$ term in the Maxwell's equations) on polarization dynamics, is non-negligible in the THz regime [27].

Theoretical and computational models for ultrafast THz excitation of ferroelectrics often fail to consider strain-polarization-photon coupling. For example, simple anharmonic oscillator models typically ignore the coupled dynamics of strain and lattice polarization [10,12,14,18], and typically employ an oversimplified one-dimensional (1D) profile of soft-mode lattice potential in the form of $aQ^2+bQ^4$, where $Q$ is the amplitude of the soft mode, thus neglecting potential 2D and 3D vibrations of the soft mode [27,28]. While molecular dynamics (MD) simulations [16,23] and dynamical phase-field simulations [21,29,30] include the coupled dynamics of the strain and lattice polarization, these both operate in the electrostatic approximation and omit the polarization-photon coupling, and hence the predicted dynamics of lattice polarization may not be sufficiently accurate due to the incapability to incorporate the polarization-current-induced damping [27] and the possible strong coupling between the lattice polarization and photons (termed as "ferron-polaritons" in [31]). Furthermore, by neglecting electrodynamics, these MD and dynamical phase-field models cannot be used to simulate the transmission of the THz wave through ferroelectric materials and other related wave phenomena such as reflection, refraction, and absorption.

Recently, a dynamical phase-field model developed by Zhuang and Hu [27], which fully couples the dynamics of strain, lattice polarization, and photons, has opened new avenues for exploring the ultrafast response of ferroelectric materials. Using this framework, Chen *et al.* investigated the transmission of a continuous THz wave across a ferroelectric $BaTiO_3$ membrane with a thickness varying from a few nanometers (nm) to a few micrometers [32], where the amplitude of the THz wave is kept low to ensure that the polarization oscillation is within the harmonic (linear) regime of the free energy landscape. Zhu *et* *al*. developed an analytical model for computing the second-order nonlinear susceptibility tensor of polarization to continuous wave (CW) THz excitation in both bulk and strained ferroelectric thin films in the framework of the Landau-Ginzburg-Devonshire (LGD) theory [33].

In this paper, we develop an LGD-based analytical model that enables predicting the transmission of a THz pulse of an arbitrary temporal waveform through a freestanding or strained ferroelectric nanomembrane. Our analytical model provides a comprehensive theoretical basis for understanding the THz transmission experiments in ferroelectrics, which often involves the use of a THz pulse. Furthermore, our calculation results suggest an approach to reversing the chirality of a circularly polarized THz pulse by harnessing the resonant polarization-photon coupling in ferroelectrics. Our results significantly advance the fundamental understanding of the interaction between THz waves and ferroelectric/polar materials, which underpin the design and development of many critical next-generation technologies ranging from THz spectroscopy to high-data-rate wireless communication to sub-THz electromechanical resonators [34].

## II. Methods

### A. Thermodynamic analysis for a ferroelectric nanomembrane uniaxially strained along any arbitrarily oriented in-plane axis

The electric Helmholtz free energy of a single-domain ferroelectric material is given by,

$$f(T, P_i, E_i, \varepsilon_{ij}) = g_0(T) + \Delta f = g_0(T) + f^{\mathrm{Landau}}(T, P_i) + f^{\mathrm{Elec}}(E_i, P_i) + f^{\mathrm{Elast}}(\varepsilon_{ij}, P_i), \quad (1)$$

where $g_0(T)$ is the Gibbs free energy density of the centrosymmetric reference phase with zero spontaneous polarization, zero electric field, and zero stress, $i,j$=1,2,3 refers to the crystal physics coordinates ($x_1$, $x_2$, $x_3$) system of this nonpolar phase. $f^{\mathrm{Landau}}(T, P_i)$ is the Landau free energy density describing the change in the Gibbs free energy for the formation of a lattice polarization $P_i$ [35] in the absence of electric and stress fields. The electrostatic and elastic energy densities $f^{\mathrm{Elec}}(E_i, P_i)$ and $f^{\mathrm{Elast}}(\varepsilon_{ij}, P_i)$ describe the interaction between the $P_i$ and the electric field $E_i$ or strain $\varepsilon_{ij}$, respectively. Here, $f^{\mathrm{Elec}}(E_i, P_i) = -0.5\kappa_0\kappa_{\mathrm{b}}E_iE_j - E_iP_i$, where $\kappa_0$ is the vacuum permittivity; $\kappa_{\mathrm{b}}$ is the background dielectric permittivity describing the contribution from the electronic polarization and vacuum [35–37]; electric field $E_i$ is the sum of the depolarization field $E_i^{\mathrm{d}}$ and the external electric field $E_i^{\mathrm{inc}}$ (the electric-field component of the incident THz pulse). For a single-domain ferroelectric nanomembrane, $E_i^{\mathrm{d}} \approx [0,0,-P_z/(\kappa_0\kappa_{\mathrm{b}})]$, where $P_z$ is the out-of-plane polarization component. The large out-of-plane component of $\mathbf{E}^{\mathrm{d}}$ would stabilize an in-plane polarization. The $f^{\mathrm{Elast}}$ is given by,

$$f^{\mathrm{Elast}}(\varepsilon_{ij}, P_i) = \frac{1}{2}c_{ijkl}(\varepsilon_{kl} - \varepsilon_{kl}^0)(\varepsilon_{ij} - \varepsilon_{ij}^0). \quad (2)$$

Here, $c_{ijkl}$ is the elastic stiffness tensor, $\varepsilon_{kl}$ is the total strain which describes the deformation of the system with respect to the nonpolar phase, $\varepsilon_{ij}^0$ is the eigenstrain describing the deformation of the system due to the nonpolar to polar phase transformation under a stress-free condition, with $\varepsilon_{ij}^0 = Q_{ijkl}P_kP_l$, where $Q_{ijkl}$ is the electrostrictive tensor. Detailed expressions for the $f^{\mathrm{Landau}}$, $c_{ijkl}$ and $Q_{ijkl}$ of the $BaTiO_3$, $SrTiO_3$, $Al_{1-x}Sc_xN$, and $LiNbO_3$ are provided in Appendix A.

The total strain $\varepsilon_{kl}$ needs to be calculated based on the mechanical boundary condition. In a freestanding nanomembrane, uniaxial tensile strains can be applied align along an arbitrarily oriented in-plane axis, denoted as $x'$. We will first calculate the total strain tensor in this rotated lab coordinate system $(x', y', z')$. and then transform it to the crystal physics system ($x_1, x_2, x_3$). This is because $c_{ijkl}$ and $Q_{ijkl}$ in $f^{\mathrm{Elast}}$ [38] along with $P_i$ in $f^{\text{Landau}}$ [39] are all defined in the ($x_1, x_2, x_3$) system. Specifically, we write the stress tensor in the $(x', y', z')$ system, $\boldsymbol{\sigma}' = \mathbf{c}'(\boldsymbol{\varepsilon}' - \boldsymbol{\varepsilon}^{0\prime})$, where $\boldsymbol{\varepsilon}' = \mathbf{R}\boldsymbol{\varepsilon}\mathbf{R}^{\mathrm{T}}$, $\boldsymbol{\varepsilon}^{0\prime} = \mathbf{R}\boldsymbol{\varepsilon}^{\mathbf{0}}\mathbf{R}^{\mathrm{T}}$, $\boldsymbol{\sigma}' = \mathbf{R}\boldsymbol{\sigma}\mathbf{R}^{\mathrm{T}}$, and $\mathbf{c}' = \mathbf{T}_{\boldsymbol{\sigma}}\mathbf{c}\mathbf{T}_{\varepsilon}^{-1}$ are the total strain, eigenstrain, stress, and the elastic stiffness tensor, respectively. $\mathbf{R}$ is the rotation matrix describing the counterclockwise rotation of the ($x$, $y$, $z$) to $(x', y', z')$ system around the surface normal ($z \| z'$) and the transformation matrix $\mathbf{T}_{\boldsymbol{\sigma}(\boldsymbol{\varepsilon})} = R_{ip}R_{jq}$. A uniaxially strained membrane is subjected to a constant strain boundary condition along the $x'$ ($\varepsilon_{x'x'}$=$\varepsilon^{\text{app}}$) and the traction-free boundary condition along both the $y'$ and $z'$ ($\sigma_{i'y'} = \sigma_{i'z'} = 0$, $i'$=$x', y', z'$) which allows for calculating the remaining five components of the total strain tensor $\boldsymbol{\varepsilon}'$. Details are provided in Appendix B. The $\boldsymbol{\varepsilon}'$ is then transformed to the ($x$, $y$, $z$) system via $\boldsymbol{\varepsilon} = \mathbf{R}^{\mathrm{T}}\boldsymbol{\varepsilon}'\mathbf{R}$, and further transformed to the ($x_1, x_2, x_3$) system for calculating the elastic energy density $f^{\text{Elast}}$ and in turn the $\Delta f$.

Minimizing the $\Delta f$ with respect to the $P_i$ under a given applied electric field and strain or stress allows us to identify the equilibrium polarization $P_i^0$ as a function of electric field and strain or stress. We first determine the $P_i^0$ in the ($x_1, x_2, x_3$) system, and then convert them to lab coordinate system ($x$, $y$, $z$) for presentation. Multiple numerical optimization methods available from the Scipy python library [40] are used, and the $P_i^0$ values leading to the lowest $\Delta f$ are recorded.

### B. Analytical model for the transmission of a THz pulse through a ferroelectric nanomembrane

The transmitted THz wave is a superposition of the incident and the radiation THz wave, i.e., $E_i$=$E_i^{\text{inc}}$+$E_i^{\text{rad}}$. For a 1D system where physical quantities vary only along the $z$ axis and the membrane thickness ($d$) is much smaller than the THz wavelength in the membrane, $E_i^{\text{rad}}$ is linearly proportional to the polarization current, i.e., $E_x^{\text{rad}} = -\frac{d}{2\kappa_0 c}\frac{\partial P_x}{\partial t}$, $E_y^{\text{rad}} = -\frac{d}{2\kappa_0 c}\frac{\partial P_y}{\partial t}$, and $E_z^{\text{rad}} = 0$, where $c$ is the speed of light in vacuum. Next, we analytically obtain $P_i(t)$ under an incident THz electric field by solving the governing equation of motion [27,30,32,33], given by,

$$\mu_{ij}\frac{\partial^2 P_j}{\partial t^2} + \gamma_{ij}\frac{\partial P_j}{\partial t} = E_i^{\text{Landau}} + E_i^{\text{Elast}} + E_i^{\text{d}} + E_i^{\text{inc}} + E_i^{\text{rad}}, i, j = x, y, z. \tag{3}$$

Here, $\mu_{ij}$ is the mass coefficient determined by the ionic mass and the Born effective charge [31] and $\gamma_{ij}$ is the phenomenological damping tensor. The effective electric fields $E_i^{\text{Landau}} = -\partial f^{\text{Landau}}(T, P_i)/\partial P_i$ and $E_i^{\text{Elast}} = -\partial f^{\text{Elast}}(\varepsilon_{ij}, P_i)/\partial P_i$. The incident THz electric field (i.e., one component of its wave packet) is given as $E_i^{\text{inc}} = E_i^{\text{inc,0}} e^{-\mathbf{i}\omega t}$. The lattice polarization is expanded as $P_i = P_i^0 + \sum_n \Delta P_i^{(n)}$ ($i$=$x$,$y$,$z$, $n$=1,2,3...$n_{\max}$), where $\Delta P_i^{(n)} = \Delta P_i^{0,(n)} e^{\mathbf{i}(-n\omega t + \varphi_i^{(n)})}$ is the harmonic component which possesses an angular frequency $n\omega$ and a phase shift $\varphi_i^{(n)}$ with respect to $E_i^{\text{inc}}$.

Using the perturbation method, which assumes the center of polarization oscillation is at $P_i^0$, Eq. (3) can be rewritten into a series of linear equations in the tensor format (details have been provided in [33]),

$$\boldsymbol{\mu}\frac{\partial^2 \Delta\mathbf{P}^{(1)}}{\partial t^2} + \boldsymbol{\gamma}^{\text{eff}}\frac{\partial \Delta\mathbf{P}^{(1)}}{\partial t} + \mathbf{K}\Delta\mathbf{P}^{(1)} = \mathbf{E}^{\text{inc}}, \tag{4a}$$

$$\boldsymbol{\mu}\frac{\partial^2 \Delta\mathbf{P}^{(2)}}{\partial t^2} + \boldsymbol{\gamma}^{\text{eff}}\frac{\partial \Delta\mathbf{P}^{(2)}}{\partial t} + \mathbf{K}\Delta\mathbf{P}^{(2)} + \mathbf{C}\Delta\mathbf{P}_{\text{II}}^{(1)} = 0, \tag{4b}$$

$$\boldsymbol{\mu}\frac{\partial^2 \Delta\mathbf{P}^{(3)}}{\partial t^2} + \boldsymbol{\gamma}^{\text{eff}}\frac{\partial \Delta\mathbf{P}^{(3)}}{\partial t} + \mathbf{K}\Delta\mathbf{P}^{(3)} + \mathbf{C}\Delta\mathbf{P}_{\text{II}}^{(1,2)} = 0. \tag{4c}$$

Here, $\boldsymbol{\mu}$=diag($\mu_{11}$,$\mu_{22}$,$\mu_{33}$) is the mass coefficient tensor, for which we assume $\mu_{ii} \equiv \mu$ as an approximation, $\boldsymbol{\gamma}^{\text{eff}} = \text{diag}\left(\gamma_{11} + \frac{d}{2\kappa_0 c}, \gamma_{22} + \frac{d}{2\kappa_0 c}, \gamma_{33}\right)$ is the effective damping coefficient tensor for the polarization oscillation which includes both the intrinsic contribution from crystal viscosity ($\gamma_{ii}$) and the extrinsic contribution from $E_i^{\text{rad}}$, $K_{ij} = -\frac{\partial^2 f}{\partial P_i \partial P_j}$ represents the local curvature in the free energy landscape, $C_{i\alpha} = \frac{1}{2}\frac{\partial^3 f}{\partial P_i \partial P_j \partial P_k}$ ($i$,$j$,$k$=1,2,3; $\alpha$=1,2,3…6), both the $\Delta\mathbf{P}_{\text{II}}^{(1)} = \Delta P_i^{(1)} \Delta P_j^{(1)}$ and $\Delta\mathbf{P}_{\text{II}}^{(1,2)} = 2\Delta P_i^{(1)} \Delta P_j^{(2)}$ are a 6×1 matrix in the summation convention. The equation of motion for the $\Delta P_i^{(n)}$ ($n$>3) can be written down in a similar fashion. Solving Eqs. (4a-c) allows for deriving $\Delta P_i^{(n)}(t)$ under an incident THz pulse $E_i^{\text{inc}}$ of arbitrary waveform. An isotropic intrinsic damping coefficient ($\gamma_{ii} \equiv \gamma$) is assumed in the calculation for simplicity. Detailed procedures are provided in Appendix C.

## III. Results and Discussion

### A. Nonlinear transmission of a broadband THz pulse through a ferroelectric nanomembrane

Let us first consider the transmission of a linearly polarized broadband THz pulse through a $(001)_{\text{pc}}$ $SrTiO_3$ (STO) membrane (pc: pseudocubic) under a 2% uniaxial strain. Although an unclamped stochiometric STO remains paraelectric down to 0 K due to quantum fluctuations, a 2% uniaxially strained $(001)_{\text{pc}}$ STO membrane displays room temperature ferroelectricity with an equilibrium polarization $P_i^0$ along the strain axis [41]. The STO membrane thickness is set to 10 nm to ensure that the polarization oscillation is spatially uniform. The temperature is set at 300 K. The incident THz electric-field pulse takes the form of $E_x^{\text{inc}}(t) = E_x^{\text{inc},0} e^{-(t-t_0)^2/2\tau^2} \cos(\omega_0(t-t_0))$, where $\omega_0/2\pi$ and $E_x^{\text{inc},0}$ defining the peak frequency and peak amplitude of the broadband THz pulse, respectively, $\tau$ determines the pulse duration, and $t_0$ determines the location of the peak. This form, which was also used in [23], mimics the experimentally recorded THz pulses [42,43].

Here, we set $\omega_0/2\pi$=1 THz, $\tau$=0.3 ps, $t_0$=1.5 ps to create a single-cycle broadband THz pulse, $E_x^{\text{inc}}(t)$. Figure 1(a) shows the waveform of $E_x^{\text{inc}}(t)$, with a peak amplitude $E_x^{\text{inc},0}$=1.88 MV/m. The broadband nature of this THz pulse can be seen from its frequency spectrum shown in Fig. 1(b). Under this condition, there is no apparent shift in the center of polarization oscillation, as can be seen from the temporal profile in Fig. 1(a) and the low spectral amplitude of $\Delta P_x(\omega=0)$ in Fig. 1(b). Therefore, the perturbation-method-based analytical model should remain valid. To confirm this, we numerically calculate the polarization dynamics induced by the same $E_x^{\text{inc}}$ via dynamical phase-field modeling (see details in [27,32]), which describe the coupled dynamics of the lattice polarization, strain, and electromagnetic (EM) waves without a priori assumptions on the features of polarization oscillation. As shown in Figs. 1(a-b), the analytical model and dynamical phase-field simulations predict almost identical profiles of $\Delta P_x(t)$ and the frequency spectrum $\Delta P_x(\omega)$, demonstrating the validity of the analytical model.

We made three main observations from the $\Delta P_x(\omega)$ plot shown in Fig. 1(b). First, the polarization oscillation is dominated by its fundamental resonance mode, at $\omega/2\pi$=$\omega_x/2\pi$=0.744 THz, where the resonant frequency $\omega_x$ is determined by the local curvature of the free energy landscape at $P_i$=$P_i^0$, with $\omega_x = \sqrt{K_{33}/\mu}$. Second, although the driving THz pulse has its peak temporal frequency at 1 THz, there is no evident peak at 1 THz in the $\Delta P_x(\omega)$. This is because the driving 1 THz component electric field has a small duration, and the dynamic linear susceptibility $\chi_{xx}^{(1)}(\omega)$ at 1 THz is significantly smaller than its on-resonance value at 0.744 THz. The $\chi_{ii}^{(1)}(\omega)$, which is related to the polarization oscillation via $\Delta P_i^{(1)} = \kappa_0 \chi_{ii}^{(1)}(\omega) E_i^{\text{inc}}(\omega)$, can be calculated as [32],

$$\chi_{ii}^{(1)}(\omega) = \frac{1}{\kappa_0} \frac{1}{\mu(\omega_i^2 - \omega^2) - \gamma_{ii}^{\text{eff}}\omega}, i = x, y. \tag{5}$$

Third, the spectral amplitude of the second-harmonic component $\Delta P_x(2\omega_x)$ is weak (~1% of the linear component), suggesting a 'weakly nonlinear' excitation. This is consistent with the fact that the polarization oscillation is still largely within the harmonic regime of the free energy landscape, as shown in Fig. 1(c).

Figure 1(d) shows the temporal profile of the transmitted THz pulse $E_x(t)$ calculated from both the analytical model and dynamical phase-field simulations. Together with the profile of $E_x^{\text{inc}}(t)$ shown in Fig. 1(a), one can see that the $E_x(t)$ is dominated by the radiation electric field $E_x^{\text{rad}}(t)$ after ~3 ps, and the peak amplitude of $E_x^{\text{rad}}(t)$ is ~0.1 MV/m. The THz transmission spectrum, represented by $E_x(\omega)$, is shown in Fig. 1(e), where significant absorption occurs at $\omega=\omega_x$ due to the resonantly excited $\Delta P_x(\omega_x)$. There is no evident absorption at $\omega=2\omega_x$ in the $E_x(\omega)$ even though $\Delta P_x(2\omega_x)$ is noticeable (c.f., Fig. 1(b)), because the amplitude of the $E_x^{\text{rad}}$ generated by the $\Delta P_i^{(2)}(t)$ is too small.

Figure 1(f-j) present similar analyses for the case of excitation by a THz pulse with a higher peak amplitude $E_x^{\text{inc},0}$=13.16 MV/m to drive the polarization oscillation further into the anharmonic regime of the free energy landscape. We note that intense THz pulses with a peak amplitude exceeding 100 MV/m [3] or 1 MV/cm, are experimentally accessible. As shown in Fig. 1(f), the center of polarization oscillation shows a negative shift, i.e., $\Delta P_x^{(2)}(0)<0$, which is attributed to the flatter free energy landscape in the side of $P<P_x^0$. This shift can be quantified as $\Delta P_x^{(2)}(0)= \chi_{xxx}^{(2)}(0)E_x^{\text{inc}}(\omega)\, E_x^{\text{inc}}(-\omega)$ [33], which describes the rectification of two THz wave components with frequencies $\omega$ and $-\omega$. By performing the inverse Fourier transform of the $\Delta P_x(\omega=0)$ component in the frequency spectrum, we obtain a $\Delta P_x^{(2)}(0)$ of -0.01 C/m$^2$. As the magnitude of $\Delta P_x^{(2)}(0)$ becomes larger, the perturbation-method-based analytical model becomes less accurate. This can be seen from the relatively large discrepancy between the analytically and numerically calculated $\Delta P_x(t)$ in Fig. 1(e). As further shown by the $\Delta P_x(\omega)$ in Fig. 1(g), the three primary peaks in the analytically calculated curve are precisely at the harmonic values ($\omega_x$, $2\omega_x$, and $3\omega_x$), whereas the three peaks in the numerically simulated $\Delta P_x(\omega)$ appear at lower frequencies due to the reduced local curvature of the free energy landscape in the side of $P<P_x^0$, as shown in Fig. 1(h). We made three observations in the transmission spectrum, $E_x(\omega)$, as shown in Fig. 1(j). First, in the spectrum calculated from dynamical phase-field simulations, the strongest absorption now shifts to a frequency that is lower than $\omega_x$, which is consistent with $\Delta P_x(\omega)$. Second, the transmission is higher over the entire frequency range, as demonstrated by the higher spectral amplitude, in the results obtained from dynamical phase-field simulations. This is consistent with the fact that $\Delta P_x(t)$ from phase-field simulations has smaller peak amplitudes than those from the analytical model [see Fig. 1(f)]. Third, higher-order (those near $2\omega_x$ and $3\omega_x$) spectral features are not evident due to the small $E_x^{\text{rad}}$ at those frequencies.

### B. Chirality reversal of a narrowband THz pulse through ferroelectric nanomembranes

We now examine how resonant polarization-photon coupling in ferroelectric nanomembranes can reverse the chirality of a narrowband, circularly polarized THz pulse. As will be elaborated below, the key is to ensure that only one of the two orthogonal components of the THz pulse (e.g., $E_y^{\text{inc}}$) resonantly excite its conjugate polarization component (e.g., $P_y$) such that the chirality of the transmitted THz pulse is determined by $E_y^{\text{rad}}$, which will have a π phase shift with respect to $E_y^{\text{inc}}$.

We start by using the archetypal $BaTiO_3$ (BTO) as an example, and assume a spatially uniform equilibrium polarization $P_i^0$ aligning along the +$x$ direction. The two components of the incident THz pulse are:

$E_x^{\rm inc}(t)$=$E_x^{\rm inc,0}e^{-(t-t_0)^2/2\tau^2}$ sin($\omega_0(t$-$t_0)$) and $E_y^{\rm inc}(t)$= $E_y^{\rm inc,0}e^{-(t-t_0)^2/2\tau^2}$ cos($\omega_0(t$-$t_0)$). The $E_x^{\rm inc}$ lags $E_y^{\rm inc}$ by a π/2 phase, leading to a left-handed chirality. $\tau$ determines both the duration and the number of cycles of the pulse. Selection of specific values for $\tau$ also depends on the value of $\omega_0$ we intend to use. In the waveform shown in Fig. 2(a), we set $\omega_0/2\pi$=1.05 THz (the same as $\omega_y$), $\tau$=5 ps and $t_0$=15 ps to create a multicycle narrowband THz pulse that peaks at $t$=15 ps. This waveform mimics the experimentally recorded circularly polarized narrowband THz pulse [8]. As shown in Fig. 2(b), $P_y$ possesses a flatter energy landscape, therefore the resonance frequency of the $P_y$ is lower (i.e., $\omega_y<\omega_x$). The peak amplitude of the incident THz pulse ($E_x^{\rm inc,0}$=$E_y^{\rm inc,0}$=19 kV/m) is four orders of magnitude smaller than the >500 kV/cm peak field used in an earlier THz excitation experiment of a single-domain $BaTiO_3$ thin film [17]. Under this condition, the THz-pulse-induced $\Delta P_i$ is dominated by the linear response ($\Delta P_i^{(1)}$).

Although the incident THz pulse is circularly polarized, the induced polarization oscillation is linearly polarized, with a minimal phase difference between $\Delta P_x^{(1)}$ and $\Delta P_y^{(1)}$, as shown in Fig. 2(c). This is because the resonantly excited (i.e., $\omega_0$=$\omega_y$) $\Delta P_y^{(1)}$ lags behind $E_y^{\rm inc}$ by a π/2 phase [32], and the non-resonantly excited ($\omega_0 \ll \omega_x$) $\Delta P_x^{(1)}$ can almost instantaneously follow $E_x^{\rm inc}$ (which also lags behind $E_y^{\rm inc}$ by a π/2 phase). The amplitudes of $\Delta P_y^{(1)}$ are two orders of magnitude larger than those of $\Delta P_x^{(1)}$, as shown in Fig. 2(c). This is because the modulus of the dynamic susceptibility $\left|\chi_{yy}^{(1)}\right|$ is two orders of magnitude larger than $\left|\chi_{xx}^{(1)}\right|$ at $\omega_0$=$\omega_y$=2π×1.05 THz, as shown in Fig. 2(d).

The transmitted electric field $E_i$ is a superposition of $E_i^{\rm inc}$ and $E_i^{\rm rad}$. The chirality of $E_i$ is represented using $\Delta\varphi^{\rm E}=\varphi_y^{\rm E}-\varphi_x^{\rm E}+\frac{\pi}{2}$, where the $\varphi_i^{\rm E}$ denotes the phase of $E_i$ with respect to $E_i^{\rm inc}$: the $E_i$ would be left-handed if 0<$\Delta\varphi^E$<π whereas right-handed if -π<$\Delta\varphi^{\rm E}$<0. As shown in Fig. 2(e), when $\omega_0$ is away from the resonance frequencies $\omega_x$ and $\omega_y$, $E_i$ remains left-handed ($\Delta\varphi^{\rm E}$>0). This is because $E_i$ is dominated by $E_i^{\rm inc}$ ($E_i^{\rm rad}$ is small under off-resonance excitation). When the $\omega_0$ is near $\omega_x$ or $\omega_y$, the chirality of $E_i$ changes from left- to right-handed ($\Delta\varphi^E$>0) after a certain period.

The mechanism of the chirality reversal can be understood from the temporal profiles of $E_i^{\rm inc}$ and $E_i^{\rm rad}$ ($i$=$x,y$) at $\omega_0$=$\omega_y$. As shown in the top panel of Fig. 2(f), $E_x^{\rm rad}$ is much smaller than $E_x^{\rm inc}$ due to the small $\Delta P_x^{(1)}$ [c.f., Fig. 1(c)]. As a result, $E_x$ is dominated by $E_x^{\rm inc}$. In contrast, the large, resonantly excited $\Delta P_y^{(1)}$ leads to a large $E_y^{\rm rad}$. As shown in the bottom panel of Fig. 2(f), the peak amplitude of $E_y^{\rm rad}$ becomes larger than that of $E_y^{\rm inc}$ after the time $t$ exceeds 18.9 ps. After then, $E_y$ would be dominated by $E_y^{\rm rad}$, which lags $E_y^{\rm inc}$ by a π phase. Overall, before 18.9 ps, $E_x$ lags $E_y$ by ~π/2, leading to a left-handed chirality (the same as $E_i^{\rm inc}$); after 18.9 ps, $E_x$ advances $E_y$ by ~π/2, resulting in a right-handed chirality.

Under a larger damping parameter $\gamma$, the chirality reversal would occur later and eventually be eliminated when $\gamma$ is too large (Appendix D). A larger $\gamma$ leads to a lower $\Delta P_i^{(1)}$, and hence a smaller $E_i^{\rm rad}$, which cannot offset $E_i^{\rm inc}$ for inducing the chirality reversal until a later stage when $E_i^{\rm inc}$ becomes smaller.

Such chirality reversal enabled by resonant lattice polarization-photon coupling can be applied to other systems including uniaxial ferroelectrics such as $Al_{1-x}Sc_xN$ and $LiNbO_3$. Consider a 10-nm-thick ferroelectric $Al_{1-x}Sc_xN$ membrane with its equilibrium polarization $P_i^0$ aligning along the $y$||$x_3$||[0001] axis, as shown in Fig. 3(a), and an incident narrowband THz pulse that has a similar waveform to that in Fig. 2(a), yet with $\tau$=1 ps, $t_0$=3 ps, and $E_x^{\rm inc,0}$=$E_y^{\rm inc,0}$=19 kV/m. Figure 3(b) shows the calculated $P_i^0$ as a function of the Sc composition $x$ and the corresponding static susceptibility $\chi_{yy}^{(1),\mathrm{dc}}$. As shown, $\chi_{yy}^{(1),\mathrm{dc}}$, which describes the response of the lattice polarization $P_y$ to a low-frequency (down to the d.c. limit) electric field $E_y^{\mathrm{inc}}$, increases substantially near the ferroelectric-to-paraelectric phase transition. This is due to the significantly reduced curvature along $y$, as shown for example for $x$=0.6 in Fig. 3(c). The large $\chi_{ii}^{(1),\mathrm{dc}}$ leads to larger $\chi_{ii}^{(1)}$ at the resonant frequencies ($\omega_0$~$\omega_i$), as shown in Fig. 3(d). Likewise, chirality reversal occurs at $\omega_0$~$\omega_i$, as shown by the temporal evolution of the $\Delta\varphi^{\mathrm{E}}$ in Fig. 3(e). For example, at $\omega_0=\omega_y=2\pi\times2.7$ THz, $E_y$ is dominated by $E_y^{\mathrm{rad}}$ (which lags $E_y^{\mathrm{inc}}$ by a π phase) after 3.5 ps, whereas $E_x$ remains virtually the same as $E_x^{\mathrm{inc}}$. As shown by the spatial trajectory of the transmitted electric-field pulse in Fig. 3(f), the long axis of the elliptical trajectory aligns along the $x$-axis before the chirality reversal, yet along the $y$-axis after the chirality reversal. The reverse is true in the case of $\omega_0=\omega_x=2\pi\times20.19$ THz. As shown in Fig. 3(g), the long axis of the spatial trajectory aligns along the $y$-axis before the chirality reversal, whereas switches to the $x$-axis after the chirality reversal. Similar chirality reversal can occur in $LiNbO_3$ (Appendix E).

### C. Strain-tunable chirality of a THz pulse transmitted through ferroelectric nanomembranes

Ferroelectric membranes provide the freedom for controlling the strain states and the corresponding ferroelectric properties. By stretching an elastically compliant substrate [44] or employing a piezoelectric strain cell [45], one can apply a uniaxial tensile strain virtually along any in-plane axis, creating many possible in-plane crystallographic orientations. In contrast to epitaxial thin films where the strains are restricted to discrete values determined by the lattice parameters of the available single crystal substrates, membranes allow for a continuous tuning of strain magnitude [46] within the elastic limit.

To demonstrate a strain-tunable chirality, we consider a 10-nm-thick $(001)_{\mathrm{pc}}$ STO membrane that is subject to a constant strain condition along the $x$-axis ($\varepsilon_{xx}=\varepsilon^{\mathrm{app}}$) and the traction-free boundary condition along the $y$- and $z$-axis (see Sec. IIA). Compared to the uniaxial $Al_{1-x}Sc_xN$ and $LiNbO_3$, it is easier to rotate the equilibrium polarization $\mathbf{P}^0$ of the multiaxial STO to other in-plane orientations by rotating the strain axis. Furthermore, the resonant frequency of the lattice polarization in the STO is highly tunable by varying the temperature and strain, thus enabling the modulation of THz pulses of a wide variety of frequencies. For example, for a $(001)_{\mathrm{pc}}$ STO membrane under 1% strain along the $x\|[100]_{\mathrm{c}}$ axis, the resonant frequency of its lattice polarization $P_x$ ($\omega_x/2\pi$) varies from 0.6 THz to 0.5 GHz as the temperature varies from 10 K to 400 K (see Appendix F for the effect of the strain and temperature on both the $\omega_x$ and $\omega_y$).

By varying $\varepsilon^{\mathrm{app}}$ from 0 to 2% and rotating the strain axis within the $xy$ plane from 0 to π/2 (described by the azimuth angle $\theta$), the calculated equilibrium polarization $P_i^0$ in the STO membrane at 50 K are shown in Fig. 4(a). We find that $\varepsilon^{\mathrm{app}}$ needs to be larger than a threshold (0.194~0.241%) to stabilize a ferroelectric phase. This threshold, which varies moderately with the angle $\theta$, is approximately equal to the stress-free strain $\varepsilon_{x'x'}^{0}$ in the rotated lab coordinate system $(x',y',z')$. When $\varepsilon^{\mathrm{app}}$ is relatively large, $\mathbf{P}^0$ rotates abruptly from $x$- to $y$-axis as $\theta$ exceeds π/4, as also shown in Fig. 4(a). When $\varepsilon^{\mathrm{app}}$ is below ~0.7%, $\mathbf{P}^0$ rotates continuously from 0 to π/2 as $\theta$ increases from 0 to π/2. The orientation of $\mathbf{P}^0$, indicated by a different azimuth angle $\Theta$, does not necessarily align along the strain axis (described by $\theta$), as shown more clearly the inset of Fig. 4(a). The continuous rotation of $\mathbf{P}^0$ can also be seen from the $\theta$-dependent $P_x^0$ and $P_y^0$ at $\varepsilon^{\mathrm{app}}$=0.25%, as shown in the top panel of Fig. 4(b). Importantly, when $\mathbf{P}^0$ is near the $[110]_{\mathrm{c}}$ direction ($P_x^0=P_y^0$), the principal components (denoted as '$p$' and '$q$') of the static susceptibility tensor are significantly enhanced, as shown in the bottom panel of Fig. 4(b). This is due to the significantly reduced local curvature in the free energy landscape induced by a uniaxial tensile strain of specific orientation, as shown in Fig. 4(c), where the orientations of the principal axes '$p$' and '$q$' are indicated. The expressions of $\chi_{\mathrm{pp}}^{(1)}(\omega)$ and $\chi_{\mathrm{qq}}^{(1)}(\omega)$ are the same as Eq. (5), except that the resonant frequencies $\omega_x$ and $\omega_y$ need to be replaced by $\omega_{\mathrm{p}}$ and $\omega_{\mathrm{q}}$, which can be calculated as $\omega_{\mathrm{p}} = \sqrt{(A_0 + B_0)/2\mu}$ and $\omega_{\mathrm{q}} = \sqrt{(A_0 - B_0)/2\mu}$, $A_0$=$K_{11} + K_{22}$ and $B_0 = \sqrt{(K_{11} + K_{22})^2 + 4K_{12}^2}$. As an example, Figure 4(d) shows the frequency-dependent moduli of $\chi_{\mathrm{pp}}^{(1)}(\omega)$ and $\chi_{\mathrm{qq}}^{(1)}(\omega)$ at $\theta$= 0.22π.

Since the chirality of an EM wave remains unchanged after the transformation of the coordinate system, we first evaluate $E_i^{\mathrm{inc}}$, $\Delta P_i^{(1)}$, $E_i^{\mathrm{rad}}$, and $E_i$ in the principal coordinate system (*p,q*), and then transform $E_i$ back to the lab coordinate system (*x,y*) to plot its spatial trajectory. Figure 4(e) shows the temporal evolution of $\Delta\varphi^E = \varphi_q^{\mathrm{E}} - \varphi_p^{\mathrm{E}} + \frac{\pi}{2}$ at different $\omega_0$ at $\theta$=0.22π at 50 K. Due to the relatively low resonance frequencies of strained STO nanomembrane at this temperature, the range of $\omega_0$ is kept below 200 GHz, which is much lower than the ranges in Fig. 2(e) and Fig. 3(e). Accordingly, we set $\tau$=30 ps and $t_0$=90 ps to ensure the duration and number of cycles are long and large enough to yield a narrowband THz pulse. The peak field amplitudes ($E_x^{\mathrm{inc},0}$=$E_y^{\mathrm{inc},0}$=19 kV/m) are kept the same. Applying a uniaxial strain along any arbitrarily oriented in-plane axis of the STO nanomembrane uniquely enables controlling both the chirality and the profile of the transmitted chiral THz pulse. For example, at $\theta$ =0.22π, a left-handed THz pulse with $\omega_0$=$\omega_{\mathrm{p}}$=2π×72 GHz will undergo a chirality reversal and possess an elliptical spatial trajectory after transmitting through the STO membrane, as shown in the left panel of Fig. 4(f). After rotating the applied strain axis to $\theta$=0.24π (by only 3.6°), the transmitted THz pulse remains left-handed with a largely circular spatial trajectory (i.e., almost the same as $E_i^{\mathrm{inc}}$), as shown in the right panel of Fig. 4(f). This is because $\left|\chi_{\mathrm{pp}}^{(1)}\right|$ and $\left|\chi_{\mathrm{qq}}^{(1)}\right|$ at 72 GHz at $\theta$=0.24π are much smaller than those at $\theta$=0.22π. As a result, $E_i^{\mathrm{rad}}$ is not large enough to modulate $E_i^{\mathrm{inc}}$ significantly.

Among the four materials studied above, we suggest that uniaxially strained STO nanomembrane to be the first choice for experimental demonstration of the predicted chirality reversal for three reasons. First, the large dynamic susceptibility $\chi_{ii}^{(1)}$ and its wide frequency window [see Fig. 4(d)] lead to a wide frequency window for the chirality reversal, as shown in Fig. 4(e). Second, the resonant frequencies of lattice polarization are highly tunable by varying the temperature and strain, further expanding the frequency window for the chirality reversal. Third, a 2% uniaxially strained STO nanomembrane possesses nm-scale 180° domains with an equilibrium polarization aligning along the strain axis [41]. Since the linear polarization oscillation $\Delta P_i^{(1)}$ produced by two domains with opposite lattice polarization (e.g., along +*x* and -*x*) should be identical, the predicted chirality reversal would occur in both the single-domain and 180° domain STO nanomembranes.

**IV. Conclusions**

We have developed an LGD-based analytical model that can be employed to predict the temporal evolution of the lattice polarization in a ferroelectric nanomembrane in response to an incident THz pulse of arbitrary waveform, as well as the concurrent THz wave transmission. Compared to previously existing analytical theories that employ a simplified 1D anharmonic oscillator model [10,12,14,18], this new analytical model provides a theoretical basis for evaluating the Landau coefficients, the full electrostrictive tensor, and the mass and damping coefficients of ferroelectrics from THz transmission experiments. This capability is important because the high-order Landau coefficients, which describe the anharmonicity of the free energy landscape, are difficult to determine from the electronics-based dielectric permittivity measurements.

This analytical model also allows for determining the parameters of the technologically important $LiNbO_3$ and the recently discovered new ferroelectric $Al_{1-x}Sc_xN$. In this work, we constructed a more comprehensive model for the LGD free energy density (as compared to [35,47–50]) by including the Landau and electrostrictive coefficients associated with field-induced polarization along the nonpolar axes. These coefficients may play an important role when applying these materials ($LiNbO_3$ and $Al_{1-x}Sc_xN$) to practical devices such as integrated electro-optical modulators [51] and sub-THz electromechanical resonators [34,52] where the spatial distribution of the driving electric fields is highly nonuniform. Moreover, the approach of harnessing the resonant lattice polarization-photon coupling in ferroelectric/polar nanomembranes to reverse the chirality of a THz pulse, along with the chirality control from a uniaxial strain applied along any arbitrarily oriented in-plane axis, provides new opportunities for THz wave modulation without relying on complex design of THz metasurfaces [53–55]. Furthermore, we expect that this chirality reversal can also occur in substrate-supported ferroelectric/polar thin films with in-plane 180º domains or single domain, such as a coherently strained $(001)_{pc}$ STO film grown on $(110)_o$ $DyScO_3$ substrate (o: orthorhombic) that has a large-area single domain after electric poling [56]. However, the presence of a thick substrate (e.g., 0.5 mm) may complicate the analysis of the transmitted THz pulse due to the wave reflection at the interfaces and wave attenuation in the substrate.

**Acknowledgements**

This work is primarily supported by the US Department of Energy, Office of Science, Basic Energy Sciences, under Award Number DE-SC0020145 as part of the Computational Materials Sciences Program (Y.Z., A.R. V.G., L.-Q.C., and J.-M.H.). A.R. also acknowledges the support of the NSF Graduate Research Fellowship Program under Grant No. DGE1255832. X. G. and J.-M.H. acknowledge the support from the National Science Foundation (NSF) under Grant No. DMR-2237884 on the dynamical phase-field simulations in this work. The dynamical phase-field simulations were performed using Bridges at the Pittsburgh Supercomputing Center through allocation TG-DMR180076 from the Advanced Cyberinfrastructure Coordination Ecosystem: Services & Support (ACCESS) program, which is supported by NSF Grants No. 2138259, No. 2138286, No. 2138307, No. 2137603, and No. 2138296. The development of free energy density for $Al_{1-x}Sc_xN$ thin film is supported by the Air Force Office of Scientific Research under award number FA9550-24-1-0159 (J.-M.H.). Partial support for manuscript preparation was provided by the Wisconsin MRSEC (DMR-2309000).

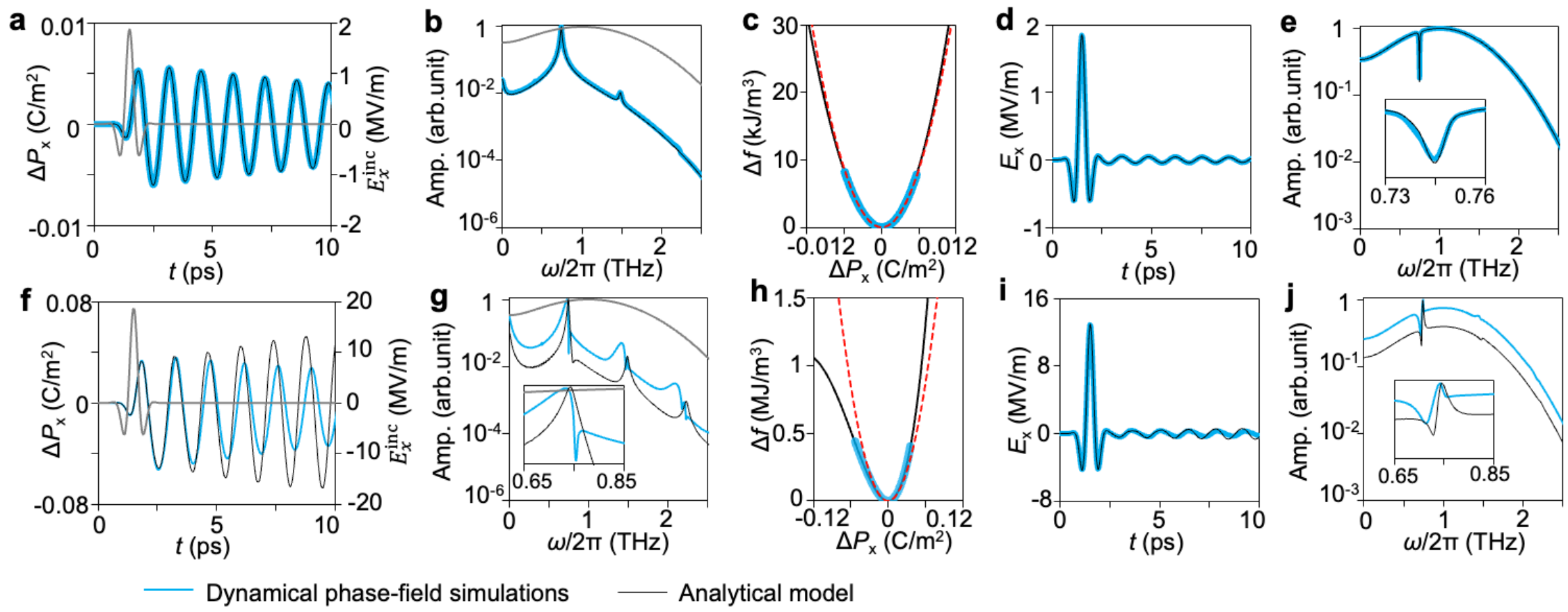

**Figure 1.** (a) Evolution of $\Delta P_x$=$P_x$-$P_x^0$, in a (001) STO membrane upon the excitation by a linearly polarized THz pulse $E_x^{\text{inc}}(t)$ at $t$=0 ps. The waveform of $E_x^{\text{inc}}(t)$ is shown on the right vertical axis. A 2% strain is applied to the 10-nm-thick STO membrane along the $x$ axis, with $x\|x_1\|[100]_c$. (b) Frequency spectra $\Delta P_x(\omega)$, obtained by Fourier transform of the $\Delta P_x(t)$ data within $t$=0-200 ps; and (gray line) $E_x^{\text{inc}}(\omega)$. The spectral amplitudes are normalized to their own largest peak amplitude. (c) 1D profile of the free energy density $\Delta f$ [see Eq. (1)] near the initial equilibrium polarization state $P_x^0$ with $P_y$=$P_z$=0 and a fixed total strain $\boldsymbol{\varepsilon}$=$\boldsymbol{\varepsilon}^0(\mathbf{P}^0)$. The thick blue line indicates the range of polarization oscillation. The dashed line is the harmonic fitting based on the local curvature at $P_x$=$P_x^0$. (d) Temporal profiles and (e) frequency spectra of transmitted THz pulse $E_x(t)$. The inset is the zoom-in figure of the absorption peak. (f-j) Data presented in a similar fashion to (a-e), yet the amplitude of $E_x^{\text{inc}}(t)$ is 10 times larger, as shown by the gray line in (f). Except (c,h), the thick blue lines show the results from the dynamical phase-field simulations, while the thin black lines indicate the results from the analytical model. The temperature is set to 300 K.

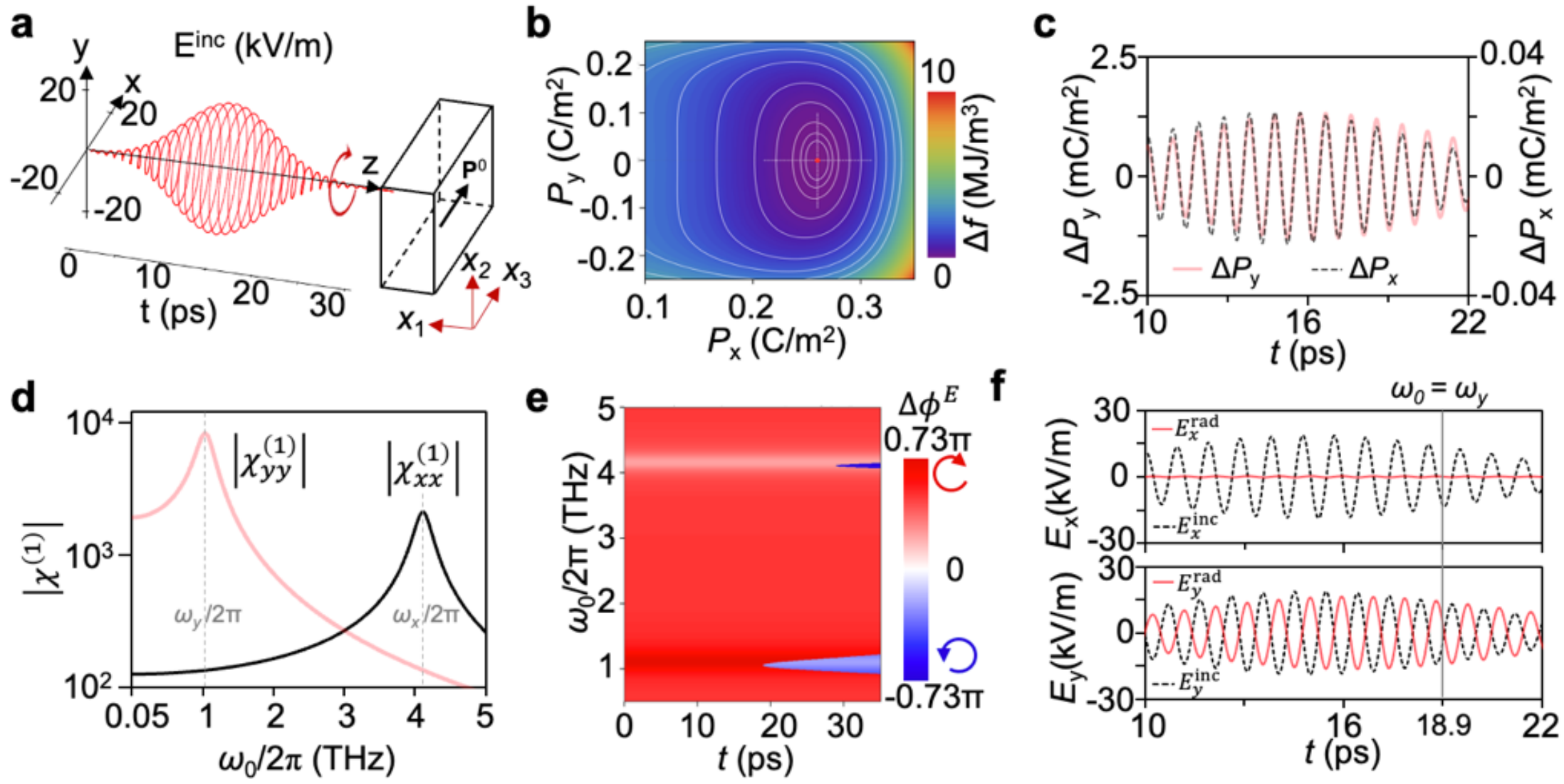

**Figure 2.** (a) Temporal profile of a left-handed narrowband chiral THz pulse that propagates along the $z$-axis of a 10-nm-thick BTO membrane that has an equilibrium polarization $P_x^0$ along the $x_3 \| x$ axis, with $x_1\|[100]_c$, $x_2\|[010]_c$, and $x_3\|[001]_c$. (b) 2D plot of the free energy density $\Delta f$ near $\mathbf{P}^0$, where $\Delta f$ is calculated based on Eq. (1) with a fixed total strain $\boldsymbol{\varepsilon}=\boldsymbol{\varepsilon}^0(\mathbf{P}^0)$. The white lines are the contour lines. The two principal axes for the dynamic susceptibility tensor $\chi_{ij}^{(1)}$ are indicated. (c) Evolution of $\Delta P_x=P_x(t)-P_x^0$ and $\Delta P_y=P_y(t)-P_y^0=P_y(t)$ within $t$=10-22 ps. $t$=0 ps refers to the moment the THz pulse reaches the membrane surface. (d) Frequency-dependent moduli of the dynamic susceptibility $\chi_{xx}^{(1)}$ and $\chi_{yy}^{(1)}$. (e) Evolution of the phase difference between the $x$ and $y$- components of the transmitted THz pulse, denoted by $\Delta\phi^{\mathrm{E}}(t)$, under different frequency $\omega_0$ of the incident narrowband THz pulse ($\tau$ and $t_0$ remain unchanged). The sign of $\Delta\phi^{\mathrm{E}}$ indicates the chirality, where the positive and negative values indicate left-handed and right-handed chirality, respectively. (f) Evolution of the $x$ and $y$- components of $E_i^{\mathrm{inc}}$ and $E_i^{\mathrm{rad}}$. The temperature is set to 300 K.

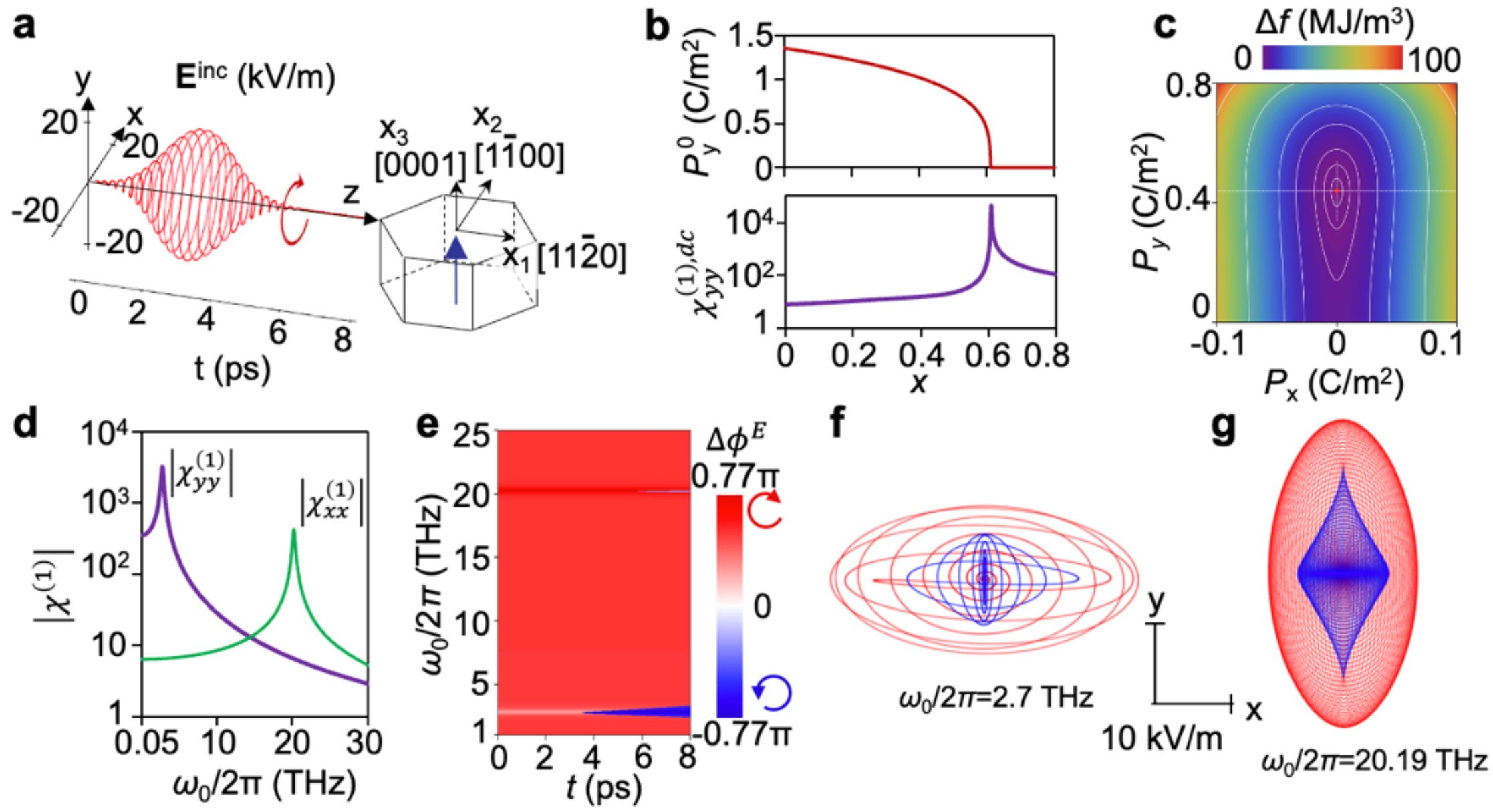

**Figure 3.** (a) Temporal profile of a left-handed narrowband THz pulse that propagates along the *z*-axis of a 10-nm-thick $Al_{1-x}Sc_xN$ nanomembrane with an equilibrium polarization $\mathbf{P}^0$ along the $x_3 \| y$ axis. (b) $P_y^0$, and $\chi_{yy}^{(1),\mathrm{dc}}$ of $Al_{1-x}Sc_xN$ as a function of *x*. (c) 2D plot of the free energy density $\Delta f$ as a function of $P_x$ and $P_y$ at *x*=0.6. Here, $\Delta f$ is calculated based Eq. (1) under the stress-free mechanical equilibrium condition (yielding $\boldsymbol{\varepsilon}=\boldsymbol{\varepsilon}^0(\mathbf{P})$), leading to $\Delta f=f^{\mathrm{Landau}}$. The white lines are the contour lines. The two principal axes of the susceptibility tensor $\chi_{ij}^{(1),\mathrm{dc}}$ are indicated. (d) Frequency-dependent moduli of the dynamic susceptibility $\chi_{xx}^{(1)}$ and $\chi_{yy}^{(1)}$. (e) Evolution of the phase difference between the *x* and *y*- components of the transmitted THz pulse, denoted by $\Delta\phi^{\mathrm{E}}(t)$, under different central frequency $\omega_0$ of the incident narrowband THz pulse ($\tau$ and $t_0$ remain unchanged). The sign of $\Delta\phi^{\mathrm{E}}$ indicates the chirality, where the positive and negative values indicate left-handed and right-handed chirality, respectively. Spatial trajectories of the transmitted THz pulse for the case of (f) $\omega_0=\omega_y$ and (g) $\omega_0=\omega_x$. The temperature is set to 300 K.

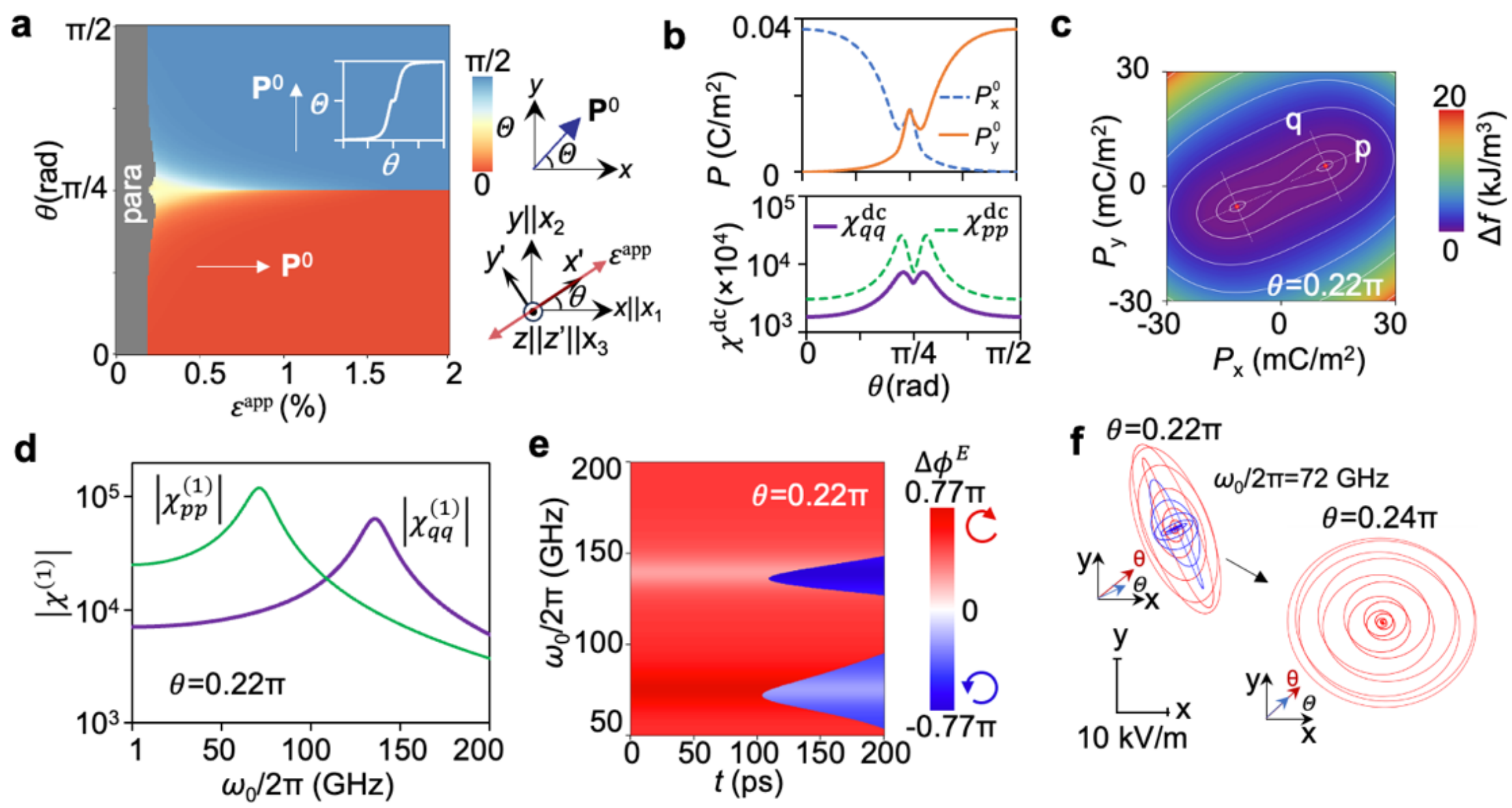

**Figure 4.** (a) Map showing the orientation of the equilibrium polarization $\mathbf{P}^0$, denoted by $\Theta$, in a (001) uniaxially strained STO nanomembrane as a function of the magnitude ($\varepsilon^{app}$) and the axis orientation ($\theta$) of the applied uniaxial strain, with $x_1||[100]_c$, $x_2||[010]_c$, and $x_3||[001]_c$. The inset shows the variation of the orientation of $\mathbf{P}^0$ with the orientation of the strain axis at $\varepsilon^{app}$=0.25%. (b) (Top) the $P_x^0$ and $P_y^0$ components of the $\mathbf{P}^0$ (note that $P_z^0$=0) and (bottom) the two principal components of the susceptibility tensor $\chi_{\mathrm{pp}}^{(1),\mathrm{dc}}$ and $\chi_{\mathrm{qq}}^{(1),\mathrm{dc}}$ as a function of the strain axis orientation $\theta$. (c) 2D density plot of the free energy density $\Delta f$ as a function of $P_x$ and $P_y$ at $\theta$=0.22π. Here, $\Delta f$ is calculated based Eq. (1) under the mixed mechanical equilibrium condition for the uniaxially strained membrane (see Sec. IIA). The white lines are the contour lines. The two principal axes ($p$ and $q$) for the susceptibility tensor $\chi_{ij}^{(1),\mathrm{dc}}$ are indicated. (d) Frequency-dependent moduli of the principal components of the dynamic susceptibility $\chi_{\mathrm{pp}}^{(1)}$ and $\chi_{\mathrm{qq}}^{(1)}$ at $\theta$=0.22π. (e) Evolution of the phase difference between the $x$ and $y$- components of the transmitted THz pulse, denoted by $\Delta\phi^{\mathrm{E}}(t)$, under different central frequency $\omega_0$ of the incident narrowband THz pulse ($\tau$ and $t_0$ remain unchanged) at $\theta$=0.22π. The sign of $\Delta\phi^{\mathrm{E}}$ indicates the chirality, where the positive and negative values indicate left-handed and right-handed chirality, respectively. (f) Spatial trajectories of the transmitted THz pulse for the case of $\omega_0/2\pi$=72 GHz at $\theta$=0.22π and $\theta$=0.24π. The temperature is set to 50 K.

**Appendix A**: The $f^{\text{Landau}}$, $c_{ijkl}$, $Q_{ijkl}$, and other parameters of the $BaTiO_3$, $SrTiO_3$, $Al_{1-x}Sc_xN$, and $LiNbO_3$

As discussed in the main paper, $f^{\text{Landau}}$ describes the change in the Gibbs free energy for the formation of an electric polarization $P_i$ in the ferroelectric (polar) phase with respect to the higher-symmetry parent nonpolar phase, in the absence of electric and mechanical fields. Within the LGD theory, the $f^{\text{Landau}}$ of a ferroelectric material with a nonpolar, centrosymmetric parent phase can be written as,

$$f^{\text{Landau}} = \alpha_{ij}P_iP_j + \alpha_{ijkl}P_iP_jP_kP_l + \alpha_{ijklmn}P_iP_jP_kP_lP_mP_n + \cdots, \quad \text{(A1)}$$

where $\alpha_{ij}$, $\alpha_{ijkl}$, $\alpha_{ijklmn}$ are the Landau coefficients under stress-free condition, with the subscripts $i, j, k, l, m, n$=1,2,3 (within the crystal physics coordinate $x_1x_2x_3$ system). $f^{\text{Landau}}$ must remain invariant under the symmetry operations, i.e.,

$$f^{\text{Landau}}(P_1, P_2, P_3) = f^{\text{Landau}}(P_1', P_2', P_3'), \quad \text{(A2)}$$

where $P_1', P_2', P_3'$ donate the polarization components in the new coordinate system ($x_1', x_2', x_3'$) transformed from the ($x_1, x_2, x_3$) system based on a given symmetry operation. This symmetry rule allows us to identify the nonzero Landau coefficients $\alpha_{ij}$, $\alpha_{ijkl}$, and $\alpha_{ijklmn}$. Taking the $LiNbO_3$ as an example, its centrosymmetric parent phase belongs to the $\bar{3}m$ point group, which entails a 120° rotation symmetry operation within the $x_1x_2$ plane: both the $x_1$ and $x_2$ axis rotate counterclockwise within their plane by an angle $\theta$=120° to $x_1'$ and $x_2'$ axis while the $x_3$ axis (the polar axis) remains unchanged ($x_3 \| x_3'$). In this case, one has **P'**=**R**.**P**, where **R** is the rotation matrix (see also Appendix B), yielding $P_1' = \cos\theta\, P_1 - \sin\theta\, P_2$, $P_2' = \sin\theta\, P_1 + \cos\theta\, P_2$, $P_3' = P_3$. Substituting the expressions of **P'** into Eq. (A1) and using Eq. (A2), one can identify the nonzero Landau coefficients. Specifically, if expanding $f^{\text{Landau}}$ up to the fourth order, only the terms $P_3^2$ and $P_1^2 + P_2^2$ remain invariant under a 120° rotation operation, i.e. $P_3^2 = {P_3'}^2$, and $P_1^2 + P_2^2 = {P_1'}^2 + {P_2'}^2$. Thus, the fourth-order Landau free energy can be constructed from a combination of $P_3^2$ and $P_1^2 + P_2^2$, i.e.,

$$f^{\text{Landau}} = \alpha_{11}(P_1^2 + P_2^2) + \alpha_{33}P_3^2 + \alpha_{1111}(P_1^2 + P_2^2)^2 + \alpha_{3333}P_3^4 + \alpha_{1133}(P_1^2 + P_2^2)P_3^2. \quad \text{(A3)}$$

The expansion up to the sixth order will be provided later in this section. For perovskite oxide ferroelectrics such as the $BaTiO_3$ and strained $SrTiO_3$, the higher-symmetry nonpolar (reference) phase is the cubic phase. An eighth-order Landau polynomial $f^{\text{Landau}}(T, P_i)$ is used for the $BaTiO_3$ and a fourth-order $f^{\text{Landau}}(T, P_i)$ is used for the $SrTiO_3$ [57], where $P_i$ is the polarization ($i$=1,2,3), with '1'$\|$[100]$_c$, '2'$\|$[010]$_c$, and '3'$\|$[001]$_c$ indicting the crystal physics coordinate system ($x_1x_2x_3$) for cubic system. Note that $P_i^0$ is used to represent the spontaneous polarization in the main paper, $P_i$ is used in the present section for simplicity.

$$\begin{aligned} f^{\text{Landau}} = {} & \alpha_1(P_1^2 + P_2^2 + P_3^2) + \alpha_{11}(P_1^4 + P_2^4 + P_3^4) + \alpha_{12}(P_1^2P_2^2 + P_2^2P_3^2 + P_1^2P_3^2) + \\ & \alpha_{111}(P_1^6 + P_2^6 + P_3^6) + \alpha_{112}[P_1^2(P_2^4 + P_3^4) + P_2^2(P_1^4 + P_3^4) + P_3^2(P_1^4 + P_2^4)] + \alpha_{123}P_1^2P_2^2P_3^2 + \\ & \alpha_{1111}(P_1^8 + P_2^8 + P_3^8) + \alpha_{1112}[P_1^6(P_2^2 + P_3^2) + P_2^6(P_1^2 + P_3^2) + P_3^6(P_1^2 + P_2^2)] + \alpha_{1122}(P_1^4P_2^4 + P_2^4P_3^4 + \\ & P_1^4P_3^4) + \alpha_{1123}(P_1^4P_2^2P_3^2 + P_1^2P_2^4P_3^2 + P_1^2P_2^2P_3^4) \end{aligned} \quad \text{(A4a)}$$

$$\begin{aligned} f^{\text{Landau}} = {} & \alpha_1(P_1^2 + P_2^2 + P_3^2) + \alpha_{11}(P_1^4 + P_2^4 + P_3^4) + \alpha_{12}(P_1^2P_2^2 + P_2^2P_3^2 + P_1^2P_3^2) + \\ & \beta_1(q_1^2 + q_2^2 + q_3^2) + \beta_{11}(q_1^4 + q_2^4 + q_3^4) + \beta_{12}(q_1^2q_2^2 + q_2^2q_3^2 + q_1^2q_3^2) - t_{11}(P_1^2q_1^2 + P_2^2q_2^2 + P_3^2q_3^2) - \\ & t_{12}[P_1^2(q_2^2 + q_3^2) + P_2^2(q_1^2 + q_3^2) + P_3^2(q_1^2 + q_2^2)] - t_{44}(P_1P_2q_1q_2 + P_2P_3q_2q_3 + P_1P_3q_1q_3) \end{aligned} \quad \text{(A4b)}$$

The elastic free energy density is given as $f^{\text{Elast}} = \frac{1}{2} c_{ijkl} e_{kl} e_{ij}$, where $e_{ij} = \varepsilon_{ij} - \varepsilon_{ij}^0$, and the elastic stiffness tensor $c_{ijkl}$ has the symmetry of the cubic parent phase for the $BaTiO_3$ and $SrTiO_3$. Accordingly, the $f^{\text{Elast}}$ can be expended as,

$$f^{\text{Elast}} = \frac{1}{2} c_{11}(e_{11}^2 + e_{22}^2 + e_{33}^2) + c_{12}(e_{11}e_{22} + e_{11}e_{33} + e_{22}e_{33}) + 2c_{44}(e_{12}^2 + e_{13}^2 + e_{23}^2) \quad \text{(A5)}$$

The eigenstrains $\varepsilon_{ij}^0 = Q_{ijkl} P_k P_l$, where the $Q_{ijkl}$ also has cubic symmetry, which can be written as,

$$\varepsilon_{11}^0 = Q_{11}P_1^2 + Q_{12}(P_2^2 + P_3^2) + \Lambda_{11}q_1^2 + \Lambda_{12}(q_2^2 + q_3^2), \quad \text{(A6a)}$$

$$\varepsilon_{22}^0 = Q_{11}P_2^2 + Q_{12}(P_1^2 + P_3^2) + \Lambda_{11}q_2^2 + \Lambda_{12}(q_1^2 + q_3^2), \quad \text{(A6b)}$$

$$\varepsilon_{33}^0 = Q_{11}P_3^2 + Q_{12}(P_1^2 + P_2^2) + \Lambda_{11}q_3^2 + \Lambda_{12}(q_1^2 + q_2^2), \quad \text{(A6c)}$$

$$\varepsilon_{23}^0 = Q_{44}P_2P_3 + \Lambda_{44}q_2q_3,\ \varepsilon_{13}^0 = Q_{44}P_1P_3 + \Lambda_{44}q_1q_3,\ \varepsilon_{12}^0 = Q_{44}P_1P_2 + \Lambda_{44}q_1q_2, \quad \text{(A6d)}$$

where $Q_{11}$, $Q_{12}$, and $Q_{44}$ are the cubic electrostrictive coefficients and $\Lambda_{11}$, $\Lambda_{12}$ and $\Lambda_{44}$ are the linear quadratic coupling coefficients between the strain and the structural order parameter [39]. At the equilibrium state, the total strain $\varepsilon_{ij}$ is determined by the mechanical boundary condition. For a stress-free (unclamped) single-domain ferroelectric membrane, $\varepsilon_{ij} = \varepsilon_{ij}^0$. The calculation of $\varepsilon_{ij}$ for a uniaxially strained single-domain ferroelectric membrane is discussed in Appendix C.

**Table 1.** List of the coefficients of the $BaTiO_3$ and $SrTiO_3$.

| Coefficients | $BaTiO_3$ | $SrTiO_3$ |
|---|---|---|
| $\alpha_1$ (N m$^2$ C$^{-2}$) | $4.124\times10^5(T\text{-}388)$ [58] | $4.05\times10^7(\coth(\frac{54}{T})\text{-}\coth(\frac{54}{30}))$ [59] |
| $\alpha_{11}$ (N m$^6$ C$^{-4}$) | $-2.097\times10^8$ [58] | $2.899\times10^9$ [56] |
| $\alpha_{12}$ (N m$^6$ C$^{-4}$) | $7.974\times10^8$ [58] | $7.766\times10^9$ [60] |
| $\alpha_{111}$ (N m$^{10}$ C$^{-6}$) | $1.294\times10^9$ [58] | 0 |
| $\alpha_{112}$ (N m$^{10}$ C$^{-6}$) | $-1.950\times10^9$ [58] | 0 |
| $\alpha_{123}$ (N m$^{10}$ C$^{-6}$) | $-2.500\times10^9$ [58] | 0 |
| $\alpha_{1111}$ (N m$^{14}$ C$^{-8}$) | $3.863\times10^{10}$ [58] | 0 |
| $\alpha_{1112}$ (N m$^{14}$ C$^{-8}$) | $2.529\times10^{10}$ [58] | 0 |
| $\alpha_{1122}$ (N m$^{14}$ C$^{-8}$) | $1.637\times10^{10}$ [58] | 0 |
| $\alpha_{1123}$ (N m$^{14}$ C$^{-8}$) | $1.367\times10^{10}$ [58] | 0 |
| $\beta_1$ (N m$^{-6}$) | 0 | $1.32\times10^{29}(\coth(\frac{145}{T})\text{-}\coth(\frac{145}{105}))$ [59] |
| $\beta_{11}$ (N m$^{-6}$) | 0 | $1.688\times10^{50}$ [59] |
| $\beta_{12}$ (N m$^{-6}$) | 0 | $3.879\times10^{50}$ [59] |
| $t_{11}$ (N m$^2$ C$^{-2}$) | 0 | $-1.902\times10^{29}$ [56] |
| $t_{12}$ (N m$^2$ C$^{-2}$) | 0 | $-1.014\times10^{29}$ [59] |
| $t_{44}$ (N m$^2$ C$^{-2}$) | 0 | 5.865×10$^{29}$ [59] |
| $c_{11}$ (GPa) | 178 [58] | 336 [59] |
| $c_{12}$ (GPa) | 96.4 [58] | 107 [59] |
| $c_{44}$ (GPa) | 122 [58] | 127 [59] |
| $Q_{11}$ (m$^4$ C$^{-2}$) | 0.1 [58] | 0.0536 [59] |
| $Q_{12}$ (m$^4$ C$^{-2}$) | -0.034 [58] | -0.0154 [59] |
| $Q_{44}$ (m$^4$ C$^{-2}$) | 0.029 [58] | 0.00472 [59] |
| $\Lambda_{11}$ (N C$^{-2}$) | 0 | 8.820×10$^{18}$ [59] |
| $\Lambda_{12}$ (N C$^{-2}$) | 0 | -7.774×10$^{18}$ [59] |
| $\Lambda_{44}$ (N C$^{-2}$) | 0 | -4.528×10$^{18}$ [59] |

For wurtzite ferroelectrics such as the $Al_{1-x}Sc_xN$, the higher-symmetry non-polar (reference) phase is typically taken as the layered hexagonal structure (point group: 6/*mmm*) [49,50,61]. A sixth-order expansion of the Landau polynomial $f^{\text{Landau}}(P_i)$ is used for the $Al_{1-x}Sc_xN$ [47,49,50],

$$f^{\text{Landau}} = \alpha_{11}(P_1^2 + P_2^2) + \alpha_{33}P_3^2 + \alpha_{1111}(P_1^2 + P_2^2)^2 + \alpha_{3333}P_3^4 + \alpha_{1133}(P_1^2 + P_2^2)P_3^2 + \alpha_{111111}(P_1^2 + P_2^2)^3 + \alpha_{333333}P_3^6 + \alpha_{111133}(P_1^2 + P_2^2)^2P_3^2 + \alpha_{113333}(P_1^2 + P_2^2)P_3^4 + \alpha_{111122}(P_1^6 - 15P_1^4P_2^2 + 15P_1^2P_2^4 - P_2^6), \quad \text{(A7)}$$

where the expansion incorporates the degeneracy of the coefficients under the 6/*mmm* symmetry, which is the point group of the centrosymmetric parent phase of the wurtzite ferroelectrics such as $Al_{1-x}Sc_xN$ [49,50,61]. The subscripts '1'||[11$\bar{2}$0] , '2'||[1$\bar{1}$00], and '3'||[0001] indicates the crystal physics coordinate system ($x_1$, $x_2$, $x_3$) in the hexagonal 6/mmm symmetry. Wurtzite $Al_{1-x}Sc_xN$ is a uniaxial ferroelectric with a spontaneous polarization aligning along the $x_3$||[0001] axis, i.e., $P_1$=$P_2$=0, $P_3$≠0. However, as a dielectric material, it can still develop nonzero polarization $P_1$ and $P_2$ upon applying (THz) electric fields along the $x_1$ and $x_2$ axis, respectively, which would in turn alter the $f^{\text{Landau}}$. To incorporate these features, we let $P_1$=$P_2$=0 in the fourth and higher-order terms of the polynomial, which are not related to the determination of the spontaneous polarization. However, we keep the quadratic term $\alpha_{11}(P_1^2 + P_2^2)$, which is related to the dielectric response along the $x_1$ and $x_2$ axis. The coefficient $\alpha_{11}$ can be derived from the experimentally measured d.c. dielectric permittivity $\kappa_{11} = \kappa_{\text{b}} + \chi_{11}^{(1),\text{dc}}$ using the relation, $\alpha_{11} = 1/\left(\kappa_0\chi_{11}^{(1),\text{dc}}\right)$, where $\kappa_0$ is the vacuum permittivity, $\kappa_{\text{b}}$ is the background permittivity and $\chi_{11}^{(1),\text{dc}}$ is the d.c. susceptibility under zero external stress.

Under these conditions, Eq. (A7) reduces to,

$$f^{\text{Landau}} = \alpha_{11}(P_1^2 + P_2^2) + \alpha_{33}P_3^2 + \alpha_{3333}P_3^4 + \alpha_{333333}P_3^6, \quad \text{(A8)}$$

As for the elastic energy density, the $c_{ijkl}$ in the $f^{\text{Elast}}$ should have the 6/mmm symmetry of the reference state with five independent components. Thus, the $f^{\text{Elast}}$ of $Al_{1-x}Sc_xN$ can be expanded as,

$$f^{\text{Elast}} = \frac{1}{2}c_{11}(e_{11}^2 + e_{22}^2) + \frac{1}{2}c_{33}e_{33}^2 + c_{12}e_{11}e_{22} + c_{13}(e_{11}e_{33} + e_{22}e_{33}) + 2c_{44}(e_{23}^2 + e_{13}^2) + 2c_{66}e_{12}^2, \quad \text{(A9)}$$

where $c_{66} = \frac{1}{2}(c_{11} - c_{12})$. The eigenstrains $\varepsilon_{ij}^0 = Q_{ijkl}P_kP_l$, where the $Q_{ijkl}$ also has the 6/*mmm* symmetry, which can be written as,

$$\varepsilon_{11}^0 = Q_{11}P_1^2 + Q_{12}P_2^2 + Q_{13}P_3^2,\ \varepsilon_{22}^0 = Q_{12}P_1^2 + Q_{11}P_2^2 + Q_{13}P_3^2,\ \varepsilon_{33}^0 = Q_{33}P_3^2 + Q_{31}(P_1^2 + P_2^2), \quad \text{(A10a)}$$

$$\varepsilon_{23}^0 = Q_{44}P_2P_3,\ \varepsilon_{13}^0 = Q_{44}P_1P_3,\ \varepsilon_{12}^0 = Q_{66}P_1P_2, \quad \text{(A10b)}$$

where $Q_{66} = Q_{11} - Q_{12}$. The $c_{ijkl}$, available Landau coefficients and $Q_{ijkl}$ of $Al_{1-x}Sc_xN$ are taken from [49], which are all functions of Sc composition *x*. The Landau coefficient $\alpha_{33}$ is also temperature-dependent, and its value at 300 K is shown in Table 2. The unavailable coefficients are estimated based on rules of crystal symmetry and elasticity, as discussed in Table 2 footnotes.

In $LiNbO_3$, the parent paraelectric phase and ferroelectric phase have a point group symmetry of $\bar{3}m$ and $3m$ (broken inversion symmetry), respectively [48]. The full expression of the sixth-order Landau free energy density for $LiNbO_3$ is the same as that for $Al_{1-x}Sc_xN$, as shown in Eq. (A7). Here, $f^{\text{Landau}}$ describes the change in the Gibbs free energy density for the formation of a ferroelectric polarization $P_i$ in the $LiNbO_3$ (3m), with respect to the parent nonpolar $\bar{3}m$, in the absence of electric and mechanical fields.

$LiNbO_3$ is a uniaxial ferroelectric, similar to the $Al_{1-x}Sc_xN$. Following [48], we use a simplified sixth-order $f^{\text{Landau}}$ for the $LiNbO_3$, i.e.,

$$f^{\text{Landau}} = \alpha_{11}(P_1^2 + P_2^2) + \alpha_{33}P_3^2 + \alpha_{1133}(P_1^2 + P_2^2)P_3^2 + \alpha_{3333}P_3^4 + \alpha_{333333}P_3^6, \quad \text{(A11)}$$

where the stress-free Landau coefficients $\alpha_{33}$, $\alpha_{3333}$, and $\alpha_{333333}$ are determined by fitting the experiment data of temperature-dependent spontaneous polarization $P_3^0$, and small-signal (**E**→0) $\chi_{33}^{(1),\text{dc}}$ collected in [35] using the relations derived for stress-free bulk crystal,

$$P_3^0 = \left(\frac{1}{3a_{333333}}\left(-a_{3333} + \sqrt{a_{3333}^2 - 3a_{33}a_{333333}}\right)\right)^{\frac{1}{2}}, \quad \text{(A12)}$$

$$\chi_{33}^{(1),\text{dc}} = \frac{1}{\kappa_0 K_{33}} = \frac{1}{\kappa_0}\left(\frac{\partial^2 f^{\text{Landau}}}{\partial P_3^2}\right)^{-1}\Bigg|_{\mathbf{P}=\mathbf{P}^0} = \frac{1}{\kappa_0\left(2a_{33} + 12a_{3333}{P_3^0}^2\right)}, \quad \text{(A13)}$$

where $a_{33}(T) = a_0\left[\coth\left(\frac{T_0}{T}\right) - \coth\left(\frac{T_0}{T_c}\right)\right]$ following [35], and the fitting gives a $T_0$=198.44 K and a Curie temperature $T_c$=1420.73 K for bulk $LiNbO_3$, the $K_{33}$ is a diagonal component of the **K** matrix (see Sec. IIB), $\mathbf{P}^0$=(0,0, $P_3^0$) denotes the initial equilibrium polarization. Moreover, the stress-free $\alpha_{11}$ and $\alpha_{1133}$ are determined by fitting the temperature-dependent small-signal(**E**→0) $\chi_{11}^{\text{dc}}$ (= $\chi_{22}^{\text{dc}}$) complied in [35] using the relations derived for stress-free bulk crystal,

$$\chi_{11}^{(1),\text{dc}} = \frac{1}{\kappa_0 K_{11}} = \frac{1}{\kappa_0}\left(\frac{\partial^2 f^{\text{Landau}}}{\partial P_1^2}\right)^{-1}\Bigg|_{\mathbf{P}=\mathbf{P}^0} = \frac{1}{\kappa_0\left(2a_{11} + 2a_{1133}{P_3^0}^2\right)}. \quad \text{(A14)}$$

We note that the elastic energy density $f^{\text{Elast}}$ was not considered in the calculation of the $\chi_{ii}^{\text{dc}}$ , *i*=1,2,3, in stress-free (fully unclamped) ferroelectric crystal, because there should be enough time for the material to reach mechanical equilibrium, yielding $\varepsilon_{ij} = \varepsilon_{ij}^0(\mathbf{P})$ (*i*,*j*=1,2,3), when the material is subjected to a low-frequency electric-field (approaching the d.c. limit). However, for calculating the $\chi_{ii}^{\text{dc}}$ of a uniaxially strained ferroelectric nanomembrane such as $SrTiO_3$ (bottom panel of Fig. 4(b)), the mechanical boundary condition, as described in Sec. IIA, needs to be employed to calculate the total strain $\varepsilon_{ij}$ and the associated $f^{\mathrm{Elast}}$. One then has $\chi_{ij}^{\mathrm{dc}} = \frac{1}{\kappa_0 K_{ij}}$, where $K_{ij} = -\frac{\partial^2 f}{\partial P_i \partial P_j}$ and $f$ is provided in Eq. (1).

Moreover, the $c_{ijkl}$ in the $f^{\mathrm{Elast}}$ should have the $\bar{3}m$ symmetry of the reference state with six independent components (different from the case of $Al_{1-x}Sc_xN$). Thus, the $f^{\mathrm{Elast}}$ of $LiNbO_3$ can be expanded as,

$$f^{\mathrm{Elast}} = \frac{1}{2}c_{11}(e_{11}^2 + e_{22}^2) + \frac{1}{2}c_{33}e_{33}^2 + c_{12}e_{11}e_{22} + c_{13}(e_{11}e_{33} + e_{22}e_{33}) + 2c_{44}(e_{23}^2 + e_{13}^2) + 2c_{66}e_{12}^2 + 2c_{14}(e_{11}e_{23} - e_{22}e_{23} + 2e_{12}e_{13}) \tag{A12}$$

where $c_{66} = \frac{1}{2}(c_{11} - c_{12})$. The eigenstrains $\varepsilon_{ij}^0 = Q_{ijkl}P_kP_l$, where the $Q_{ijkl}$ also has the $\bar{3}m$ symmetry, which be written as,

$$\varepsilon_{11}^0 = Q_{11}P_1^2 + Q_{12}P_2^2 + Q_{13}P_3^2 + Q_{14}P_2P_3,\ \varepsilon_{22}^0 = Q_{12}P_1^2 + Q_{11}P_2^2 + Q_{13}P_3^2 - Q_{14}P_2P_3, \tag{A13a}$$

$$\varepsilon_{33}^0 = Q_{31}P_1^2 + Q_{31}P_2^2 + Q_{33}P_3^2,\ \varepsilon_{23}^0 = Q_{41}P_1^2 - Q_{41}P_2^2 + Q_{44}P_2P_3, \tag{A13b}$$

$$\varepsilon_{13}^0 = Q_{44}P_1P_3 + 2Q_{41}P_1P_2,\ \varepsilon_{12}^0 = Q_{14}P_1P_3 + Q_{66}P_1P_2, \tag{A13c}$$

where $Q_{66} = Q_{11} - Q_{12}$. The $c_{ijkl}$ and available $Q_{ijkl}$ of $LiNbO_3$ are taken from [48]. Unavailable $Q_{ijkl}$ are estimated based on rules of crystal symmetry and elasticity (see Table 2 footnotes).

**Table 2**. List of the coefficients of the $Al_{1-x}Sc_xN$ and $LiNbO_3$.

| Coefficients | $Al_{1-x}Sc_xN$ | $LiNbO_3$ |
|---|---|---|
| $\alpha_{11}$ (N m$^2$ C$^{-2}$) | $8.618\times10^9$ (a) | $3.224\times10^8$ |
| $\alpha_{33}$ (N m$^2$ C$^{-2}$) | $2.75189\times10^9(x-0.6115)$ [49] | $1.174\times10^8(\coth(\frac{198.44}{T})-\coth(\frac{198.44}{1420.73}))$ |
| $\alpha_{1133}$ (N m$^6$ C$^{-4}$) | 0 | $5.804\times10^8$ |
| $\alpha_{3333}$ (N m$^6$ C$^{-4}$) | $0.03725775\times10^9$ [49] | $2.421\times10^8$ |
| $\alpha_{333333}$ (N m$^{10}$ C$^{-6}$) | $0.15476817\times10^9$ [49] | $3.392\times10^8$ |
| $c_{11}$ (GPa) | $285.12x+396(1-x)-238.39x(1-x)$ [49] | 198.86 [48] |
| $c_{12}$ (GPa) | $180.75x+137(1-x)+11.23x(1-x)$ [49] | 54.67 [48] |
| $c_{13}$ (GPa) | $141.7x+108(1-x)+51.59x(1-x)$ [49] | 67.26 [48] |
| $c_{14}$ (GPa) | 0 | 7.83 [48] |
| $c_{33}$ (GPa) | $-155.17x+373(1-x)+95.49x(1-x)$ [49] | 233.7 [48] |
| $c_{44}$ (GPa) | $176.44x+116(1-x)-158.8x(1-x)$ [49] | 59.85 [48] |
| $c_{66} = \frac{1}{2}(c_{11} - c_{12})$ (GPa) | $52.19x+129.5(1-x)-124.81x(1-x)$ [49] | 72.095 [48] |
| $Q_{13}$ (m$^4$ C$^{-2}$) | $-0.006068027-0.02272466x+0.0831565x^2-0.2269113x^3$ (b) | 0.00166466 [48] |
| $Q_{14}$ (m$^4$ C$^{-2}$) | 0 | 0.00621181 [48] |
| $Q_{33}$ (m$^4$ C$^{-2}$) | 0.0243286+0.11958864$x$-0.4443864$x^2$+1.24416$x^3$ [(b)] | -0.00856204 [48] |
| $Q_{31}$ (m$^4$ C$^{-2}$) | -0.006068027-0.02272466$x$+0.0831565$x^2$-0.2269113$x^3$[(c)] | 0.00166466[(c)] |
| $Q_{41}$ (m$^4$ C$^{-2}$) | 0 | 0.00621181[(c)] |
| $Q_{11}$ (m$^4$ C$^{-2}$) | 0.014597-0.07175$x$-0.26663$x^2$+0.74649$x^3$ [(c)] | 0.005137224[(c)] |
| $Q_{12}$ (m$^4$ C$^{-2}$) | -0.00291944 - 0.0143506 $x$ + 0.0533264 $x^2$- 0.149299 $x^3$ [(c)] | -0.00256861[(c)] |
| $Q_{44}$ (m$^4$ C$^{-2}$) | 0.0072986 + 0.0358766 $x$ - 0.133316 $x^2$ + 0.373248 $x^3$ [(c)] | -0.0334318 [48] |

(a) The $\alpha_{11}$ of $Al_{1-x}Sc_xN$ was estimated based on the resonant frequency of the polarization $P_1$ along the $x_1$||[11$\bar{2}$0] axis, denoted as $\omega_1$, based on the relationship $\omega_1 = \sqrt{K_{11}/\mu}$, where $\mu$ is the mass coefficient. The $\omega_1$ is equivalent to the resonant frequency of the *E1* transverse optical (TO) phonon mode in the wurtzite $Al_{1-x}Sc_xN$. Due to lack of data, we set the $\omega_1$ to the value of pure AlN, $\omega_1$=2π×20.07 THz (wavenumber: 669 cm$^{-1}$) [62]. The estimation of the $\mu$ will be discussed below.

(b) The present expressions of $Q_{13}$ and $Q_{33}$ are multiplied by a factor of 0.7 and 1.2, respectively, from the expressions provided in Ref. [49]. This slight tuning was performed to ensure the calculated Sc composition ($x$) dependence of the dynamic dielectric susceptibility $\chi_{ij}^{(1)}(\omega)$ is physically reasonable (i.e., having a Lorentzian line shape as described in Eq. (5)) in the range of $x$ = 0.5 - 0.8.

(c) We could not find the values of $Q_{31}$, $Q_{11}$, $Q_{12}$, and $Q_{44}$ of $Al_{1-x}Sc_xN$ in the literature. We estimate them as, $Q_{31}\approx Q_{13}$, $Q_{11}\approx 0.6Q_{33}$, $Q_{12}\approx -0.5Q_{11}$, $Q_{44}\approx 0.3Q_{33}$. Similarly, the values of $Q_{11}$, $Q_{12}$, $Q_{31}$, and $Q_{41}$ of $LiNbO_3$ remains elusive. We estimate them as $Q_{11}\approx -0.6Q_{33}$, $Q_{12}\approx -0.5Q_{11}$, $Q_{31}\approx Q_{13}$, $Q_{41}\approx Q_{14}$. The rationales for these estimates are provided below.

It is worth pointing out that the $Q_{31}$ and $Q_{13}$ in $Al_{1-x}Sc_xN$ should not be equal by crystal symmetry [63]. Specifically, the $Q_{31}$ quantifies the contribution of the $P_1^2$ to the eigenstrain $\varepsilon_{33}^0$ and $Q_{13}$ quantifies the contribution of the $P_3^2$ to the strain $\varepsilon_{11}^0$, as indicated in Eq. (A13a-b) Under a 60° $x_1$-$x_2$ plane rotation operation $R$, required by the hexagonal symmetry of $Al_{1-x}Sc_xN$, the strain tensor $\varepsilon_{ij}$ and the polarization product $P_kP_l$ transform as, $\varepsilon_{ij} = R_{im}R_{jn}\varepsilon_{mn}$, $P_kP_l = R_{kp}R_{lq}P_pP_q$, and the transformed terms $\varepsilon_{mn}$, $(P_pP_q)$ must equal to the original terms $\varepsilon_{ij}$, $(P_kP_l)$, respectively. This leads to the constraint on the electrostrictive tensor, $Q_{ijkl} \equiv Q_{mnpq} = R_{jn}^{-1}R_{im}^{-1}Q_{ijkl}R_{kp}R_{lq}$. Compared to the four-rank stiffness tensor $c_{ijkl}$, which arises from the second derivative of the elastic energy density, $f^{\text{Elas}} = \frac{1}{2}c_{ijkl}\varepsilon_{ij}\varepsilon_{kl}$, and therefore symmetrical, $c_{ijkl} = c_{klij}$, the electrostrictive tensor only need to satisfy the aforementioned constraint $Q_{ijkl} \equiv Q_{mnpq}$ required by the hexagonal symmetry, but not required to satisfy $Q_{ijkl} = Q_{klij}$. This is summarized in [62] yet without explanation. Likewise, we have $Q_{31}\neq Q_{13}$ and the $Q_{41}\neq Q_{14}$ in $LiNbO_3$.

In both the $Al_{1-x}Sc_xN$ and $LiNbO_3$, $Q_{11}$ should be positive but smaller in magnitude compare to $Q_{33}$ because increasing $|P_1|$ will lead to lattice expansion along the $x_1$ axis ($\varepsilon_{11}^0$>0) and because the electrostrictive response in the $x_1x_2$ plane is weaker than those along the polar axis $x_3$; $Q_{12}$ is negative because increasing $|P_2|$ will lead to compression along the $x_1$ axis ($\varepsilon_{11}^0$<0); it is also reasonable to consider that the $|Q_{12}| < |Q_{11}|$,similarly to 'cubic' ferroelectrics (see Table S1); Likewise, $Q_{44}$ is positive because a positive shear deformation within the $x_1x_3$ or $x_2x_3$ plane ($\varepsilon_{13}^0$>0 or $\varepsilon_{23}^0$>0) should occur when $P_1P_3$>0 or $P_2P_3$>0 according to its crystal structure (see Eq. (A10b)). However, one would expect $|Q_{44}| < |Q_{33}|$ because the normal deformation along the polar axis $x_3$ should be more significant than the shear deformation in both the $Al_{1-x}Sc_xN$ and $LiNbO_3$. We note that these assumptions provide a qualitative understanding of the electrostrictive coefficients in these systems and future experiments or first-principles calculations may provide a more complete understanding of the coupling between stress and polarization.

The diagonal mass coefficient tensor $\mu_{ii}$ can be calculated as $\mu_{ii}=MV/(eZ_{ii})^2$ [14], $i$=1,2,3, where $M$ is the soft-mode reduced mass, $V$ is the unit cell volume, $e$ is the elementary charge, and $Z_{ii}$ is the Born effective charge tensor of the soft mode [64]. For simplicity, we assume an isotropic Born effective charge tensor, and hence an isotropic mass coefficient $\mu$. Specifically, one has $\mu$ = 1.35×10$^{-18}$ J m s$^2$ C$^{-2}$ for $BaTiO_3$ [31], $\mu$ = 22×10$^{-18}$ J m s$^2$ C$^{-2}$ for $SrTiO_3$ [65], $\mu$ = 1.81×10$^{-18}$ J m s$^2$ C$^{-2}$ for $LiNbO_3$ [31]. For $Al_{1-x}Sc_xN$, $\mu$ = 1.0866×10$^{-18}$ J m s$^2$ C$^{-2}$. This value is estimated based on the resonant frequency of the $P_3$ along the $x_3$||[0001] axis, denoted as $\omega_3$, based on the relationship $\omega_3 = \sqrt{K_{33}/\mu}$. The $\omega_3$ is equivalent to the resonant frequency of the A1 transverse optical (TO) phonon mode in the wurtzite $Al_{1-x}Sc_xN$. Here we set the $\omega_3$ to the value of pure AlN, with $\omega_3$=2π×18.29 THz (wavenumber $\upsilon$=610 cm$^{-1}$) [62].

**Appendix B**: Calculation of the total strain in ferroelectric nanomembranes under a uniaxial strain applied along any arbitrarily oriented in-plane axis

The analysis of the total strain in a ferroelectric membrane under a uniaxial tensile strain applied along an arbitrary in-plane axis involves the use of three coordinate systems: crystal physics coordinate system ($x_1$, $x_2, x_3$), lab coordinate system ($x, y, z$), and the rotated lab coordinate system under the applied in-plain strain ($x', y', z'$). The relationship among these coordinate systems has been illustrated in the figures and discussed in the main paper. For the case shown in Fig. 4a, the coordinates in the ($x, y, z$) and ($x', y', z'$) system can be converted via the relation,

$$\begin{bmatrix} x' \\ y' \\ z' \end{bmatrix} = \boldsymbol{R}.\begin{bmatrix} x \\ y \\ z \end{bmatrix}, \boldsymbol{R} = \begin{bmatrix} \cos\theta & \sin\theta & 0 \\ -\sin\theta & \cos\theta & 0 \\ 0 & 0 & 1 \end{bmatrix} \tag{B1}$$

The stress-free (unclamped) ferroelectric/polar nanomembrane exhibits a spontaneous strain $\varepsilon^0_{x'x'}$, which is calculated as the $\varepsilon^0_{x'x'} = (x_1^1 - x_1^0)/x_1^0$ using the higher-symmetry nonpolar phase as the reference. The strain $\varepsilon^{\text{app}}$, which is applied along the $x'$ axis, is typically calculated via $\varepsilon^{\text{app}} = (x_1^2 - x_1^1)/x_1^1$ using the stress-free ferroelectric phase as the reference. Here the $x_1^2$, $x_1^1$, and $x_1^0$ are the coordinates along the $x'$ axis. The total strain $\varepsilon_{x'x'}$, which needs to be calculated using the nonpolar phase as the reference, is given by,

$$\varepsilon_{x'x'} = (x_1^2 - x_1^0)/x_1^0 = (\varepsilon^0_{x'x'} + 1)(\varepsilon^{\text{app}} + 1) - 1. \tag{B2}$$

The other components of the total strain can be calculated by considering the traction-free boundary condition, given as,

$$\sigma_{x'y'} = \sigma_{y'y'} = \sigma_{x'z'} = \sigma_{y'z'} = \sigma_{z'z'} = 0. \tag{B3}$$

We note that $\varepsilon^0_{x'x'}$=0 in the case of $SrTiO_3$ because a stress-free $SrTiO_3$ is nonpolar with no macroscopic spontaneous polarization. Based on Eq. (B3), we expand the stress-strain relation in the ($x', y', z'$) system,

$$\begin{pmatrix} \sigma_{x'x'} \\ \sigma_{y'y'} \\ \sigma_{z'z'} \\ \sigma_{y'z'} \\ \sigma_{x'z'} \\ \sigma_{x'y'} \end{pmatrix} = \begin{pmatrix} \sigma_{x'x'} \\ 0 \\ 0 \\ 0 \\ 0 \\ 0 \end{pmatrix} = \mathbf{c}' \begin{pmatrix} \varepsilon_{x'x'} - \varepsilon^0_{x'x'} \\ \varepsilon_{y'y'} - \varepsilon^0_{y'y'} \\ \varepsilon_{z'z'} - \varepsilon^0_{z'z'} \\ 2\left(\varepsilon_{y'z'} - \varepsilon^0_{y'z'}\right) \\ 2\left(\varepsilon_{x'z'} - \varepsilon^0_{x'z'}\right) \\ 2\left(\varepsilon_{x'y'} - \varepsilon^0_{x'y'}\right) \end{pmatrix}, \tag{B4}$$

where $\mathbf{c}'$ denotes the elastic stiffness tensor in the ($x', y', z'$) system which is obtained by transforming the stiffness tensor $\mathbf{c}$ in the ($x, y, z$) system. The eigenstrain components can be calculated via $\boldsymbol{\varepsilon}^{0\prime} = \mathbf{R}.\boldsymbol{\varepsilon}^0.\mathbf{R}^{\text{T}}$, where the $\boldsymbol{\varepsilon}^0$ are the eigenstrains in the ($x, y, z$) system. We note that the expressions of the eigenstrains and the c in Appendix A are both defined in the ($x_1, x_2, x_3$) system. Therefore, additional transformation will be needed to convert them to the ($x, y, z$) system. The total strain components $\boldsymbol{\varepsilon}'$ (other than the $\varepsilon_{x'x'}$) can be expanded using the relations $\boldsymbol{\varepsilon}' = \mathbf{R}.\boldsymbol{\varepsilon}.\mathbf{R}^{\text{T}}$ and $\boldsymbol{\sigma}' = \mathbf{R}.\boldsymbol{\sigma}.\mathbf{R}^{\text{T}}$, i.e.,

$$\begin{pmatrix} \varepsilon_{x'x'} \\ \varepsilon_{y'y'} \\ \varepsilon_{z'z'} \\ 2\varepsilon_{y'z'} \\ 2\varepsilon_{x'z'} \\ 2\varepsilon_{x'y'} \end{pmatrix} = \mathbf{T}_{\boldsymbol{\varepsilon}}. \begin{pmatrix} \varepsilon_{xx} \\ \varepsilon_{yy} \\ \varepsilon_{zz} \\ 2\varepsilon_{yz} \\ 2\varepsilon_{xz} \\ 2\varepsilon_{xy} \end{pmatrix}, \mathbf{T}_{\boldsymbol{\varepsilon}} = \begin{pmatrix} \cos^2\theta & \sin^2\theta & 0 & 0 & 0 & -\cos\theta\sin\theta \\ \sin^2\theta & \cos^2\theta & 0 & 0 & 0 & \cos\theta\sin\theta \\ 0 & 0 & 1 & 0 & 0 & 0 \\ 0 & 0 & 0 & \cos\theta & \sin\theta & 0 \\ 0 & 0 & 0 & -\sin\theta & \cos\theta & 0 \\ 2\cos\theta\sin\theta & -2\cos\theta\sin\theta & 0 & 0 & 0 & \cos^2\theta - \sin^2\theta \end{pmatrix}, \quad \text{(B5a)}$$

$$\begin{pmatrix} \sigma_{x'x'} \\ \sigma_{y'y'} \\ \sigma_{z'z'} \\ \sigma_{y'z'} \\ \sigma_{x'z'} \\ \sigma_{x'y'} \end{pmatrix} = \mathbf{T}_{\boldsymbol{\sigma}}. \begin{pmatrix} \sigma_{xx} \\ \sigma_{yy} \\ \sigma_{zz} \\ \sigma_{yz} \\ \sigma_{xz} \\ \sigma_{xy} \end{pmatrix}, \mathbf{T}_{\boldsymbol{\sigma}} = \begin{pmatrix} \cos^2\theta & \sin^2\theta & 0 & 0 & 0 & -2\cos\theta\sin\theta \\ \sin^2\theta & \cos^2\theta & 0 & 0 & 0 & 2\cos\theta\sin\theta \\ 0 & 0 & 1 & 0 & 0 & 0 \\ 0 & 0 & 0 & \cos\theta & \sin\theta & 0 \\ 0 & 0 & 0 & -\sin\theta & \cos\theta & 0 \\ \cos\theta\sin\theta & -\cos\theta\sin\theta & 0 & 0 & 0 & \cos^2\theta - \sin^2\theta \end{pmatrix}, \quad \text{(B5b)}$$

The $\mathbf{T}_{\boldsymbol{\varepsilon}}$ and $\mathbf{T}_{\boldsymbol{\sigma}}$ matrix allows us to calculate the $\mathbf{c}'$ via $\mathbf{c}' = \mathbf{T}_{\boldsymbol{\sigma}}.\mathbf{c}.\mathbf{T}_{\boldsymbol{\varepsilon}}^{-1}$. Substituting Eqs. (B5-a,b) into Eq. (B4) yields the following six linear equations which allow for solving the $\sigma_{x'x'}$ and $\varepsilon_{ij}$, $ij \neq x'x'$,

$$\begin{pmatrix} \sigma_{x'x'} \\ \varepsilon_{y'y'} \\ \varepsilon_{z'z'} \\ 2\varepsilon_{y'z'} \\ 2\varepsilon_{x'z'} \\ 2\varepsilon_{x'y'} \end{pmatrix} = \mathbf{A}^{-1}.\mathbf{T}_{\boldsymbol{\sigma}}.\mathbf{c}.\mathbf{T}_{\boldsymbol{\varepsilon}}^{-1}. \begin{pmatrix} (1+\varepsilon^{\mathrm{app}})(1+\varepsilon^{0}_{x'x'}) - 1 \\ 0 \\ 0 \\ 0 \\ 0 \\ 0 \end{pmatrix} - \mathbf{A}^{-1}.\mathbf{T}_{\boldsymbol{\sigma}}.\mathbf{C}. \begin{pmatrix} \varepsilon^{0}_{x'x'} \\ \varepsilon^{0}_{y'y'} \\ \varepsilon^{0}_{z'z'} \\ 2\varepsilon^{0}_{y'z'} \\ 2\varepsilon^{0}_{x'z'} \\ 2\varepsilon^{0}_{x'y'} \end{pmatrix}, \quad \text{(B6)}$$

Where $\mathbf{A} = \begin{bmatrix} 1 & 0 \\ 0 & \mathbf{c}'_{2:6,2:6} \end{bmatrix}$, and $\mathbf{c}'_{2:6,2:6}$ denotes the submatrix consisting of rows and columns 2 through 6 of the full 6 × 6 elastic stiffness matrix $\mathbf{c}'$. One can then obtain the total strain in the $(x, y, z)$ system via,

$$\begin{pmatrix} \varepsilon_{xx} \\ \varepsilon_{yy} \\ \varepsilon_{zz} \\ 2\varepsilon_{yz} \\ 2\varepsilon_{xz} \\ 2\varepsilon_{xy} \end{pmatrix} = \mathbf{T}_{\boldsymbol{\varepsilon}}^{-1}. \begin{pmatrix} \varepsilon_{x'x'} \\ \varepsilon_{y'y'} \\ \varepsilon_{z'z'} \\ 2\varepsilon_{y'z'} \\ 2\varepsilon_{x'z'} \\ 2\varepsilon_{x'y'} \end{pmatrix}, \quad \text{(B7)}$$

Finally, the total strain tensor in the $(x_1, x_2, x_3)$ system can be obtained via,

$$\begin{pmatrix} \varepsilon_{11} \\ \varepsilon_{22} \\ \varepsilon_{33} \\ 2\varepsilon_{23} \\ 2\varepsilon_{13} \\ 2\varepsilon_{12} \end{pmatrix} = \mathbf{T}_{\boldsymbol{\varepsilon}}^{\mathrm{lab}\rightarrow\mathrm{cryst}}.\mathbf{T}_{\boldsymbol{\varepsilon}}^{-1}. \begin{pmatrix} \varepsilon_{x'x'} \\ \varepsilon_{y'y'} \\ \varepsilon_{z'z'} \\ 2\varepsilon_{y'z'} \\ 2\varepsilon_{x'z'} \\ 2\varepsilon_{x'y'} \end{pmatrix}, \quad \text{(B8)}$$

Where $\mathbf{T}_{\boldsymbol{\varepsilon}}^{\mathrm{lab}\rightarrow\mathrm{cryst}}$ is the transformation matrix between the lab coordinate system $(x, y, z)$ and the crystal physics coordinate system $(x_1, x_2, x_3)$. For the case where $x \| x_1$, $y \| x_2$, and $z \| x_3$, $\mathbf{T}_{\boldsymbol{\varepsilon}}^{\mathrm{lab}\rightarrow\mathrm{cryst}}$ equals to the identity matrix.

**Appendix C**: Derivation of the analytical formulate of $\Delta P_i^{(n)}(t)$ in ferroelectric membranes

To derive $\Delta\mathbf{P}^{(1)}(t)$, we rewrite Eq. (4a) in Sec. IIB into the following,

$$\frac{\partial\mathbf{P}^{(1)*}(t)}{\partial t}=\mathbf{A}\mathbf{P}^{(1)*}(t)+\mathbf{F}^{(1)}(t), \tag{C1}$$

where $\mathbf{P}^{(1)*}(t)=\left(\Delta P_x^{(1)},\Delta P_y^{(1)},\Delta P_z^{(1)},\frac{\partial\Delta P_x^{(1)}}{\partial t},\frac{\partial\Delta P_y^{(1)}}{\partial t},\frac{\partial\Delta P_z^{(1)}}{\partial t}\right)^{\mathrm{T}}$, $\mathbf{F}^{(1)}(t)=\left(0,0,0,E_x^{\mathrm{inc}},E_y^{\mathrm{inc}},E_z^{\mathrm{inc}}\right)^{\mathrm{T}}$, and the coefficient matrix can be written as,

$$\mathbf{A}=\begin{pmatrix}0 & \mathbf{I}\\ -\mathbf{K}/\mu & -\boldsymbol{\gamma}^{\mathrm{eff}}/\mu\end{pmatrix}, \tag{C2}$$

where $\mathbf{I}$ is a 3×3 identity matrix. Using the initial conditions of $\Delta P_i^{(1)}(t=0)=0$ and $\frac{\partial\Delta P_i^{(1)}}{\partial t}(t=0)=0$, the time-domain solution to Eq. (C1) can be written as,

$$\mathbf{P}^{(1)*}(t)=\sum_{i=1}^{6}\frac{\boldsymbol{v}_i}{\boldsymbol{w}_i^{\mathrm{T}}\boldsymbol{v}_i}\int_0^t e^{\lambda_i(t-\zeta)}\,\boldsymbol{w}_i^{\mathrm{T}}\mathbf{F}^{(1)}(\zeta)\mathrm{d}\zeta, \tag{C3}$$

where $\lambda_i$, $\boldsymbol{w}_i^{\mathrm{T}}$ and $\boldsymbol{v}_i$ (i=1,2,3,4,5,6) are the eigenvalues, left eigenvectors, and right eigenvectors of the coefficient matrix $\mathbf{A}$, respectively.

Likewise, $\Delta\mathbf{P}^{(2)}(t)$ and $\Delta\mathbf{P}^{(3)}(t)$ can be derived by rewriting Eqs. (4b) and (4c) into the following forms:

$$\frac{\partial\mathbf{P}^{(2)*}(t)}{\partial t}=\mathbf{A}\mathbf{P}^{(2)*}(t)+\mathbf{F}^{(2)}(\mathrm{t}), \tag{C4a}$$

$$\frac{\partial\mathbf{P}^{(3)*}(t)}{\partial t}=\mathbf{A}\mathbf{P}^{(3)*}(t)+\mathbf{F}^{(3)}(\mathrm{t}). \tag{C4b}$$

Here, $\mathbf{P}^{(n)*}(t)=\left(\Delta P_x^{(n)},\Delta P_y^{(n)},\Delta P_z^{(n)},\frac{\partial\Delta P_x^{(n)}}{\partial t},\frac{\partial\Delta P_y^{(n)}}{\partial t},\frac{\partial\Delta P_z^{(n)}}{\partial t}\right)^{\mathrm{T}}$, $n$=2 or 3, $\mathbf{F}^{(2)}(\mathrm{t})=(0,0,0,-\left(\mathbf{C}\Delta\mathbf{P}_{\mathrm{II}}^{(1)}\right)_1,$ $-\left(\mathbf{C}\Delta\mathbf{P}_{\mathrm{II}}^{(1)}\right)_2,-\left(\mathbf{C}\Delta\mathbf{P}_{\mathrm{II}}^{(1)}\right)_3)^{\mathrm{T}}$, and $\mathbf{F}^{(2)}(\mathrm{t})=(0,0,0,-\left(\mathbf{C}\Delta\mathbf{P}_{\mathrm{II}}^{(1,2)}\right)_1,-\left(\mathbf{C}\Delta\mathbf{P}_{\mathrm{II}}^{(1,2)}\right)_2,-\left(\mathbf{C}\Delta\mathbf{P}_{\mathrm{II}}^{(1,2)}\right)_3)^{\mathrm{T}}$, where the subscripts '1','2','3' indicates the first, second, and third component of the 3×1 matrix $\mathbf{C}\Delta\mathbf{P}_{\mathrm{II}}^{(1)}$ or $\mathbf{C}\Delta\mathbf{P}_{\mathrm{II}}^{(1,2)}$. Using the initial conditions of $\Delta P_i^{(n)}(t=0)=0$ and $\frac{\partial\Delta P_i^{(n)}}{\partial t}(t=0)=0$, $n$=2 or 3, the time-domain solution to Eq. (C4a) and (C4b) can be written as,

$$\mathbf{P}^{(2)*}(t)=\sum_{i=1}^{6}\frac{\boldsymbol{v}_i}{\boldsymbol{w}_i^{\mathrm{T}}\boldsymbol{v}_i}\int_0^t e^{\lambda_i(t-\zeta)}\,\boldsymbol{w}_i^{\mathrm{T}}\mathbf{F}^{(2)}(\zeta)\mathrm{d}\zeta, \tag{C5a}$$

$$\mathbf{P}^{(3)*}(t)=\sum_{i=1}^{6}\frac{\boldsymbol{v}_i}{\boldsymbol{w}_i^{\mathrm{T}}\boldsymbol{v}_i}\int_0^t e^{\lambda_i(t-\zeta)}\,\boldsymbol{w}_i^{\mathrm{T}}\mathbf{F}^{(2)}(\zeta)\mathrm{d}\zeta. \tag{C5b}$$

Below we show the expanded analytical expressions of $\Delta P_i^{(1)}(t)$ as an example. In both the tetragonal ($BaTiO_3$, strained $SrTiO_3$) and the hexagonal ($LiNbO_3$ and $Al_{1-x}Sc_xN$) ferroelectric membranes, we assume that each axis of the lab coordinate system ($x$, $y$, $z$) is aligned with one of the axes of the crystal physics coordinate system ($x_1, x_2, x_3$). Specifically, the out-of-plane direction corresponds to the z-axis. In this case, the off-diagonal elements in the matrix $\mathbf{K}$ are equal to zero, i.e. $K_{ij} = 0$, $i \neq j$. The explicit expression of $\lambda_i$, $\boldsymbol{w}_i^{\mathrm{T}}$, and $\boldsymbol{v}_i$ (i=1,2,3,4,5,6) in Eqs. (C5a-b) are given as,

$$\lambda_1 = \lambda_2^* = -\lambda_x + \mathrm{i}\omega_x,\ \lambda_3 = \lambda_4^* = -\lambda_y + \mathrm{i}\omega_y,\ \lambda_5 = \lambda_6^* = -\lambda_z + \mathrm{i}\omega_z, \tag{C6a}$$

$$\boldsymbol{v}_1 = \boldsymbol{v}_2^* = \left(\frac{\mu}{K_{xx}}(-\lambda_x - \mathrm{i}\omega_x), 0,0,1,0,0\right)^{\mathrm{T}}, \tag{C6b}$$

$$\boldsymbol{v}_3 = \boldsymbol{v}_4^* = \left(0, \frac{\mu}{K_{yy}}(-\lambda_y - \mathrm{i}\omega_y), 0,0,1,0\right)^{\mathrm{T}}, \tag{C6c}$$

$$\boldsymbol{v}_5 = \boldsymbol{v}_6^* = \left(0,0, \frac{\mu}{K_{zz}}(-\lambda_z - \mathrm{i}\omega_z), 0,0,1\right)^{\mathrm{T}}, \tag{C6d}$$

$$\boldsymbol{w}_1 = \boldsymbol{w}_2^* = (\lambda_x + \mathrm{i}\omega_x, 0,0,1,0,0)^{\mathrm{T}}, \tag{C6e}$$

$$\boldsymbol{w}_3 = \boldsymbol{w}_4^* = \left(0, \lambda_y + \mathrm{i}\omega_y, 0,0,1,0\right)^{\mathrm{T}}, \tag{C6f}$$

$$\boldsymbol{w}_5 = \boldsymbol{w}_6^* = (0,0, \lambda_z + \mathrm{i}\omega_z, 0,0,1)^{\mathrm{T}}, \tag{C6g}$$

$$\lambda_x = \frac{\gamma_{xx}^{\mathrm{eff}}}{2\mu},\ \lambda_y = \frac{\gamma_{yy}^{\mathrm{eff}}}{2\mu},\ \lambda_z = \frac{\gamma_{zz}}{2\mu}, \tag{C6h}$$

$$\omega_x = \frac{1}{2\mu}\sqrt{4\mu K_{xx} - {\gamma_{xx}^{\mathrm{eff}}}^2},\ \omega_y = \frac{1}{2\mu}\sqrt{4\mu K_{yy} - {\gamma_{yy}^{\mathrm{eff}}}^2},\ \omega_z = \frac{1}{2\mu}\sqrt{4\mu K_{zz} - {\gamma_{zz}}^2}, \tag{C6i}$$

According to the eigenvectors, we found that the motions of the three components in $\Delta P_i^{(1)}(t)$, $i$=$x$,$y$,$z$, are independent of each other. We assume the incident field $\boldsymbol{E}^{\mathrm{inc}}(t) = \left(E_x^{\mathrm{inc}}(t), E_y^{\mathrm{inc}}(t), E_z^{\mathrm{inc}}(t)\right)$, the solution of the $\Delta P_i^{(1)}$ can be expanded as,

$$\Delta P_x^{(1)} = e^{-\lambda_x t}\cos\omega_x t\left(-\frac{1}{\mu\omega_x}\int_0^t e^{\lambda_x\zeta}\sin\omega_x\zeta\, E_x^{\mathrm{inc}}(\zeta)\mathrm{d}\zeta\right) + e^{-\lambda_x t}\sin\omega_x t\left(+\frac{1}{\mu\omega_x}\int_0^t e^{\lambda_x\zeta}\cos\omega_x\zeta\, E_x^{\mathrm{inc}}(\zeta)\mathrm{d}\zeta\right), \tag{C7a}$$

$$\Delta P_y^{(1)} = e^{-\lambda_y t}\cos\omega_y t\left(-\frac{1}{\mu\omega_y}\int_0^t e^{\lambda_y\zeta}\sin\omega_y\zeta\, E_y^{\mathrm{inc}}(\zeta)\mathrm{d}\zeta\right) + e^{-\lambda_y t}\sin\omega_y t\left(+\frac{1}{\mu\omega_y}\int_0^t e^{\lambda_y\zeta}\cos\omega_y\zeta\, E_y^{\mathrm{inc}}(\zeta)\mathrm{d}\zeta\right), \tag{C7b}$$

$$\Delta P_x^{(1)} = e^{-\lambda_z t}\cos\omega_z t\left(-\frac{1}{\mu\omega_z}\int_0^t e^{\lambda_z\zeta}\sin\omega_z\zeta\, E_z^{\mathrm{inc}}(\zeta)\mathrm{d}\zeta\right) + e^{-\lambda_z t}\sin\omega_z t\left(+\frac{1}{\mu\omega_z}\int_0^t e^{\lambda_z\zeta}\cos\omega_z\zeta\, E_z^{\mathrm{inc}}(\zeta)\mathrm{d}\zeta\right). \tag{C7c}$$

The expressions in Eqs. (C7a-c) are mathematically equivalent to the established solution of an anharmonic oscillator model that describes the motion of a soft mode amplitude ($Q$) driven by a pulsed electric field [66], where the soft-mode lattice potential takes a simpler form of $aQ^2+bQ^4$.

**Appendix D**: Influence of the intrinsic polarization damping parameter on the chirality reversal

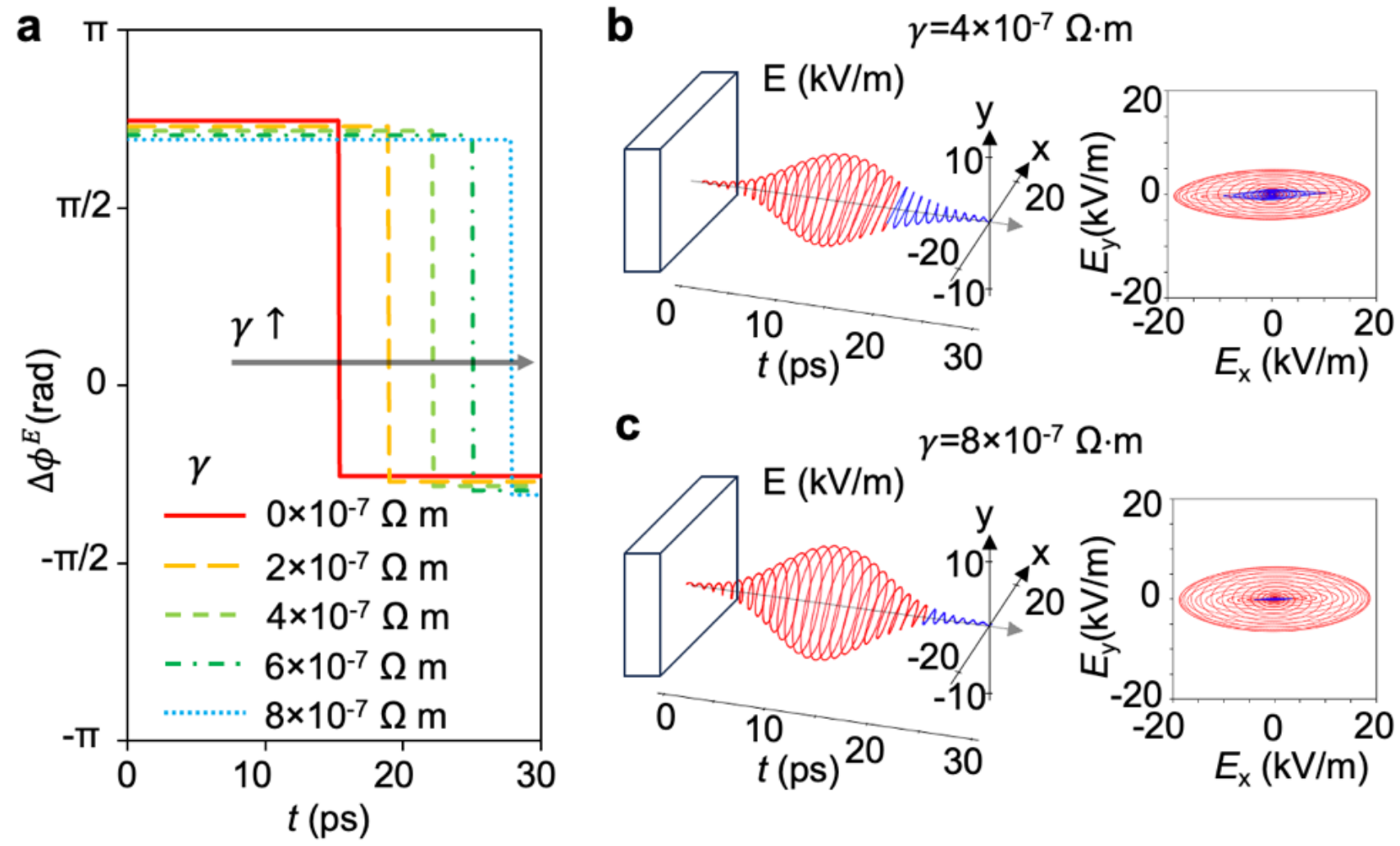

**Figure 5.** (a) The evolution of the phase difference $\Delta\phi^E$ for the same condition as those in Fig. 2(e), with $\omega_0/2\pi=\omega_y/2\pi$=1.05 THz but various values for the damping coefficient $\gamma$. (b, c) (Left) temporal and (right) spatial trajectory of the transmitted THz pulse under $\gamma$=4×10$^{-7}$ Ω·m and 8×10$^{-7}$ Ω·m, where the chirality reversal happens at $t$=22.1 ps and 27.9 ps, respectively.

**Appendix E**: Chiral THz transmission through a $LiNbO_3$ nanomembrane

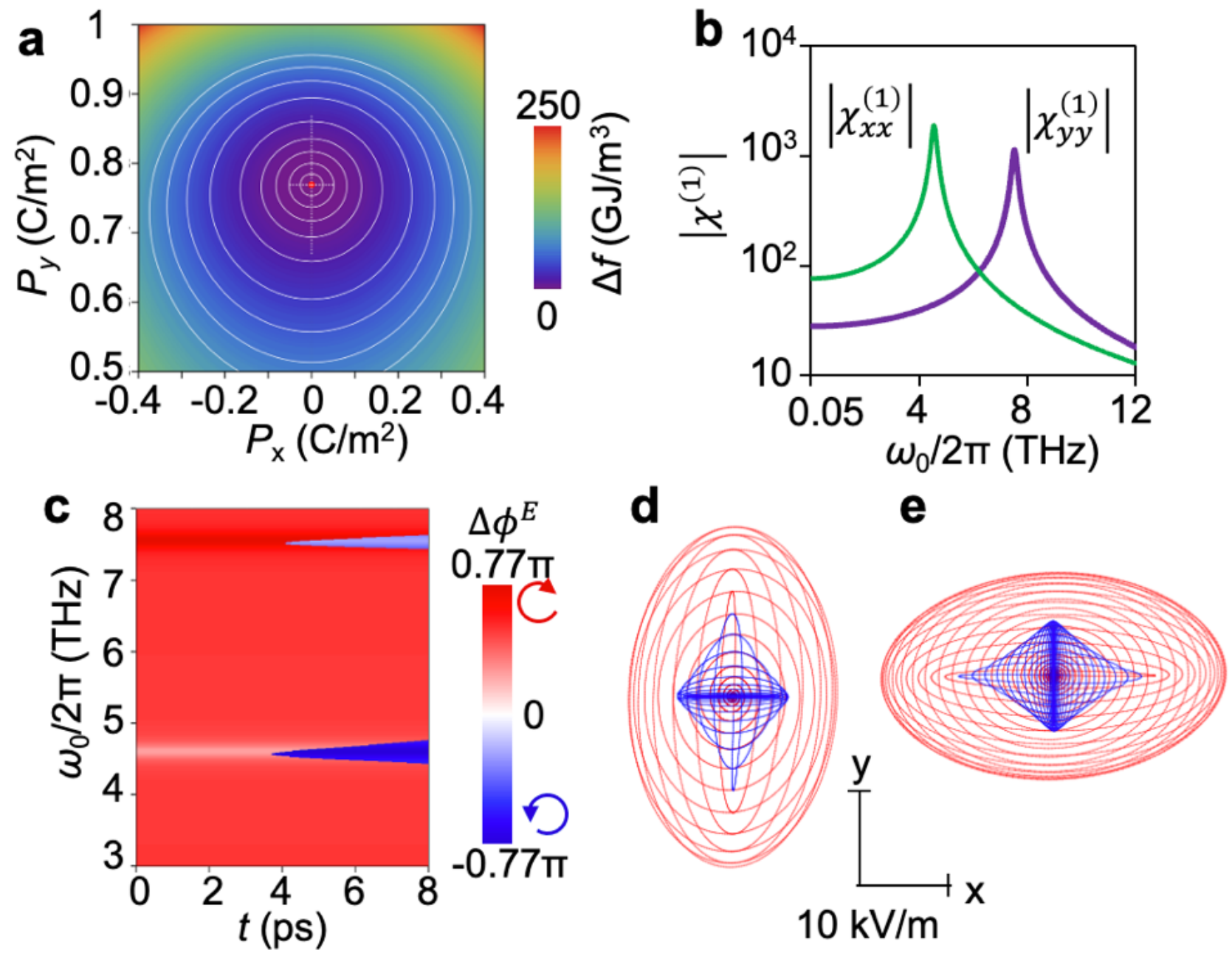

**Figure 6.** (a) 2D plot of the free energy density $\Delta f$ near the $\mathbf{P}^0$. Here, $\Delta f$ is calculated based on Eq. (2) under the stress-free mechanical equilibrium condition (yielding $\boldsymbol{\varepsilon}=\boldsymbol{\varepsilon}^0(\mathbf{P})$), leading to $\Delta f=f^{\text{Landau}}$. The white lines are the contour lines. The two principal axes for the susceptibility tensor $\chi_{ij}^{(1),\text{dc}}$ are indicated. (b) Frequency-dependent moduli of the dynamic susceptibility $\chi_{xx}^{(1)}$ and $\chi_{yy}^{(1)}$. (c) Evolution of the phase difference between the $x$ and $y$- components of the transmitted THz pulse, denoted by $\Delta\phi^{\text{E}}(t)$, under different central frequency $\omega_0$ of the incident THz pulse ($\tau$ and $t_0$ remain unchanged). The sign of $\Delta\phi^{\text{E}}$ indicates the chirality, where the positive and negative values indicate left-handed and right-handed chirality, respectively. Spatial trajectories of the transmitted THz pulse for the case of (d) $\omega_0=\omega_x$ and (e) $\omega_0=\omega_y$. The temperature is set to 300 K. The values of and $\tau$=1 ps and $t_0$=3 ps used in the incident THz pulse are same as those used for $Al_{0.4}Sc_{0.6}N$ in Fig. 3.

**Appendix F**: The effect of the applied strain $\varepsilon^{\text{app}}$ and temperature on the $\omega_x$ and $\omega_y$ of a (001) STO nanomembrane uniaxially strained along the $x||[100]_c$ axis.

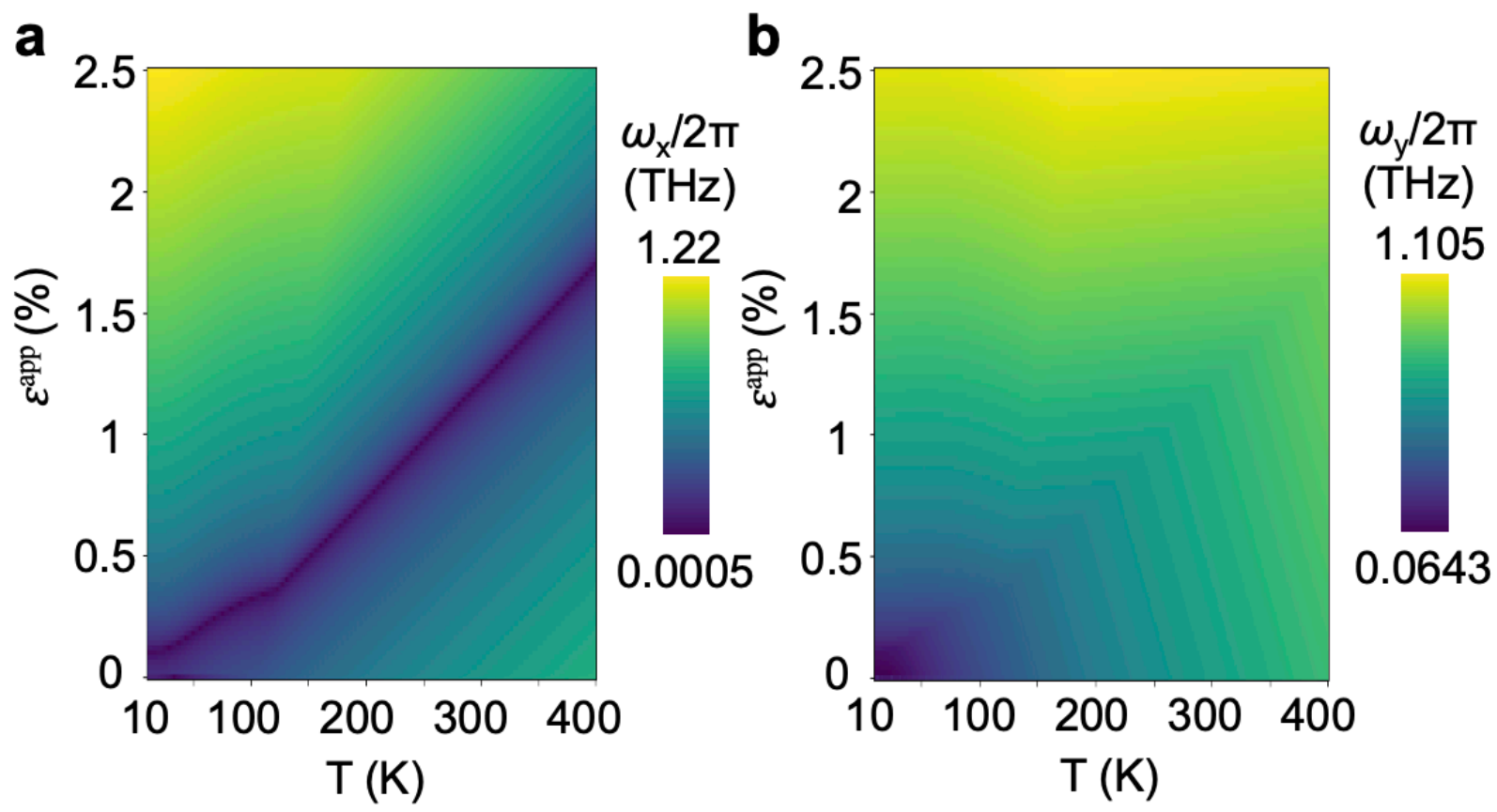

**Figure 7.** Resonant frequency of the polarization (a) $P_x$ and (b) $P_y$, denoted as $\omega_x/2\pi$ and $\omega_x/2\pi$, respectively, as a function of the magnitude of the applied strain $\varepsilon^{\text{app}}$ and temperature in the (001) STO nanomembrane. Note that $x||x_1||[100]_c$, $y||x_2||[010]_c$, and $z||x_3||[001]_c$.